\documentclass[10pt,a4paper]{article}
\usepackage{graphpap,epsfig,amssymb,graphicx,textcomp,xcolor}
\begin{document}

\begin{center}
{\Large {\bf Critical behaviours of  anisotropic XY ferromagnet in the presence of random field}}\end{center}

\vskip 1cm

\begin{center}{\it Olivia Mallick$^1$ and Muktish Acharyya$^{2,*}$}\\
{\it Department of Physics, Presidency University,}\\
{\it 86/1 College Street, Kolkata-700073, INDIA}\\
{$^1$E-mail:olivia.rs@presiuniv.ac.in}\\
{$^2$E-mail:muktish.physics@presiuniv.ac.in}\end{center}

\vskip 1cm

\noindent {\bf Abstract:} The anisotropic XY ferromagnet has been studied by Monte Carlo simulation in a three dimensional
simple cubic lattice. The increase in critical temperature (ferro-para transition) has been noticed with increasing 
the strength of anisotropy. The effects of random fields (both with full circular symmetry and in angular window) on the critical temperature  are investigated systematically in the anisotropic XY ferromagnet in three dimensions. Reduction of the critical temperature of anisotropic XY ferromagnet has been observed in the presence of random field. The compensating field
(the required amount of field which preserves the critical temperature
for {\it isotropic} XY ferromagnet) has been studied as a function of the strength of anisotropy. The compensating field was found to depend linearly on the strength of anisotropy. We have also studied the effects of random field confined in the angular window and observed the reduction of the critical temperature with increase of the angular extension. The critical behaviours are formalized by the usual finite size analysis and the estimation of critical exponents
for the susceptibility and the specific heat.

\vskip 3cm


\vskip 2cm

\noindent $^*$ Corresponding author
\newpage

\noindent {\bf {I. Introduction:}}

In condensed matter physics and statistical physics, phase transition\cite{stanley} is widely studied subject. The ordered phase is mainly characterised by long range order or correlations.
However, the special kind of phase transition in the SO(2) symmetric planar ferromagnet has drawn\cite{plischke,chaikin} intense attention of researchers in last 
few decades. The phase in the absence of long range ferromagnetic 
ordering is the peculiarity of such kind of phase transition\cite{kosterlitz}. 

The critical behaviours of isotropic (SO(2) symmetric) XY ferromagnet has been studied
\cite{betts} by
exact high-temperature series expansion method and predicted the
critical temperature ${{kT_c} \over {J}}=4.84\pm0.06$ and estimated the critical exponents. The three dimensional XY ferromagnet has
been investigated \cite{campostrini,hasenbusch} later by Monte Carlo simulation to estimate the critical temperature
($T_c = 2.206...$) and specified the XY universality class.

How the anisotropy (which breaks SO(2) symmetry) plays a role in the critical behaviour of XY ferromagnets ? This important question has been addressed by a number of researches in recent past. The
anisotropic effect was introduced\cite{shenoy} by the difference in the inter-plane/intra-plane coupling of three dimensional XY ferromagnet. They observed the dependence of the critical temperature on the strength/amount of the anisotropy. In the vanishingly small interplanar coupling they also found the 
Kosterlitz-Thouless limit of phase transition. The one-dimensional quasiperiodic anisotropic XY model was found\cite{satija} to exhibit ordered and disordered phases with cantor spectra. Interestingly, it was also observed that at a particular point, the quasiperiodic anisotropic XY model in one dimension, exhibited a pointlike spectrum with localized states.
The {\it quantum} anisotropic XY ferromagnetic system was studied
\cite{ma} to analyse the anisotropy dependent critical temperature. The critical temperature has been found\cite{ma} to increase with increasing the strength of anisotropy. Recently,
the classical anisotropic XY ferromagnet has been studied\cite{olivia} extensively by Monte Carlo simulation. For constant anisotropy, the critical temperature was found to increase as the strength of the anisotropy increased. Here, the critical behaviour, of statistically distributed anisotropy, has been investigated systematically. But, interestingly, the critical temperature has been found to decrease\cite{olivia} as the width of the distribution of the anisotropy increased.

Let us briefly mention here some other relevant studies on the XY model. The XY model with antinematic interaction has been studied
\cite{zuko1}. The geometrically frustrated generalised XY model has been studied\cite{zuko2} to have new kind of ordered phase. The
antiferromagnetic XY model with higher order interaction has been 
investigated\cite{zuko3} and the phase diagram been drawn. Recently,
the Monte Carlo method has been employed to study\cite{erol} the various phases of layered XY antiferromagnet.


Can one realise the role of the anisotropy experimentally ? Let
us briefly mention some related experimental studies to realise
the effects of anisotropy. The spin transport and the Kondo effect are seriously affected by the anisotropy. In the Kondo effect, the
differential conductance shows a single peak. However, the anisotropy could split the single peak into two distinct peaks
in single $Co$ atom bound on the top of a $Cu$ atom of 
the $Cu_2N$ surface.

The anisotropy creates the degenerate ground-state in the large
spin atoms. These degenerate ground-states are connected by the spin flip of a screening electron. This fact is responsible for the emergence of Kondo resonance.
The magnetic
anisotropy also plays a major role in determining\cite{otte} how the Kondo resonance evolves in a magnetic
field: the resonance peak splits at a rate that is strongly direction dependent. The anisotropy governs the directionality here. The anisotropy plays a significant role in the spin transport also.  

The role of magnetic anisotropy in spin-filter junction may be referred here\cite{chopdekar}. The magneto
- transport is largely governed by the magnetic anisotropy at the interface or junction. The
system is fabricated LSMO/chromite/$Fe_3O_4$ junctions where the chromite barrier layer, either
$CoCr_2O_4 (CCO)$ or $MnCr_2O_4(MCO)$, is isostructural with 
$Fe_3O_4$. These ultrathin chromite
layers exhibited normal ferromagnetic behaviours below their bulk Curie temperature ($T_c$) and
proximity-induced ferromagnetism due to $Fe_3O_4$ above their bulk $T_c$ , thus giving rise to an effective
spin-filter junction. Although both chromite compounds form a normal spinel structure with all
$Cr^{3+}$ ions in the octahedral sites, the magnetic anisotropy of the two compounds are opposite in
sign consequently giving rise to junction magnetoresistance values more than an order of magnitude
higher in CCO junctions compared to that of MCO junctions. The
Kosterlitz-Thouless phase transition has been experimentally observed\cite{murthy} in a system of ultracold Fermi gas.

The XY ferromagnet in the presence of magnetic field has been investigated\cite{rosa,gouveat}.
Since the random field seriously affects the critical behaviour 
of Ising ferromagnets, it would be an interesting question,
what would be the effect of random magnetic field on the critical behaviours of XY ferromagnet. Recently, the domain growth and aging are studied\cite{puri} in random field XY (RFXY) model. The 
quasi long range ordered (QLRO) phase in the limit of low disorder (random field) has been predicted 
\cite{fisher,feldman} in the form of {\it Bragg glass} phase in random field XY model.  A similar
topological phase transition to a pinned vortex-free phase at a
nonzero critical field strength in three dimensional random field XY systems has been predicted
from numerical studies\cite{fisch}. The vortex-glass phase is also predicted \cite{garanin} from 
the numerical studies on three-dimensional random field XY model at zero temperature.

However, the results of these studies are challenged\cite{tissier, tarjus} by claiming that the lower critical dimension  for the
quasi long range ordered  phase is 3.9 (obtained from functional renormalization group study), i.e., there is no QLRO in d = 2, 3 RFXY
systems !! The range of lower critical dimension in RFXY model is supported\cite{kutay} recently.

But the study on the effect of random field in the {\it anisotropic} XY ferromagnet is missing in the literature. 
The ordered phase in the presence of random magnetic field can only be found\cite{tarjus}in the dimension 3.9 of {\it isotropic} (SO(2) symmetric) XY ferromagnet. However,
for the broken SO(2) symmetric  anisotropic XY ferromagnet, it would not be surprising to expect the order-disorder transition in three dimensional XY model, in the presence of random magnetic field. 
The random magnetic field would play the role of random disorder which may lead to the ferro-para
phase transition in anisotropic XY model in three dimensions. The anisotropy would increase the
critical temperature. In contrary, the random field would tend to reduce the critical temperature. 
{\it So, can one find any suitably adjusted pair of values
of random field and the anisotropy, which may yield the critical temperature for isotropic (SO(2)
symmetric) XY ferromagnet in three dimensions ?}
This motivated us to study the critical behaviour of three dimensional {\it anisotropic} XY model in the presence of random field. Here, we have considered that the magnitude of the random field is fixed but the direction if random
(i) within 0 and $2\pi$ and (ii) within angular window which does not have full circular symmetry. We employed Monte Carlo simulation method to study the critical behaviours of three dimensional anisotropic XY ferromagnet in the presence of random field. The paper is organised as follows: the model is introduced and the Monte Carlo simulation scheme is described in the next section. The numerical results and analysis are reported in section-III, the paper ends with
summary and concluding remarks in section-IV.

\newpage

\noindent {\bf {II.Model and Simulation :}}

The  classical XY model describes a system of spins with continuous symmetry (SO(2)). The spins are represented by two-dimensional vectors that lie in a plane. The exchange interaction is characterised by a sum over all pairs of nearest neighbour spins and weighted by an interaction strength $J$. The anisotropy compels the spins to be aligned along a preferred direction and is controlled by the parameter $\gamma$. The Hamiltonian of such system in presence of  a random field is represented by

\begin{equation}
\mathcal{H}= -J \sum_{<i,j>} [(1+\gamma)S_i^xS_j^x + (1-\gamma)S_i^yS_j^y] -\sum_{i} \vec{h_i}.\vec{S_i}
\end{equation}

Here ${\vec S_{i}} (=cos \theta_i,sin \theta_i)$ is the two-dimensional vector (spin)  with unit length ($|{\vec S}|=1$) specified by angle $\theta_i$ which can take any values (classical) between 0 to $2\pi$. $J>0$ is the ferromagnetic interaction strength between the neighbouring spins. Here, $\gamma$ is the anisotropy parameter or the strength of anisotropy. The random field  $\vec h_{i} (=h cos \phi_i, h sin \phi_i)$ in each (denoted by i-th site) lattice site represents the quenched disorder, described by field strength $h=|\vec h_i|$ and angle variable $\phi_i$. The randomness of the  field is governed by the random directions only (magnitude of field 
$h$ is constant). The average (over the lattice) value of the random field is zero ($<\vec h_i>=0$). The magnitude of field is measured in the unit of $J$. The anisotropy (dimensionless) term is taken in such a way that positive values of 
$\gamma$ promote alignment along X-direction while negative values favour perpendicular (to the X-direction) alignment. Here, the first summation represents the summation over distinct nearest neighbour lattice sites.
This corresponds to energy due to spin-spin interaction. The second summation represents the energy of interaction with external magnetic field.

We study the equilibrium phase transition of a three dimensional (simple cubic of size $L=20$)  anisotropic XY model in presence of such random field. The periodic boundary conditions are imposed in all three directions. Initially, the system is at a high-temperature paramagnetic phase with random initial orientations of spins, i.e. $<{\vec S_i}>=0$. At a finite temperature $T$ (measured in units of $J/{k}$, where $k$ is the Boltzmann constant), a lattice site is randomly selected from the system with an initial configuration $\theta_i(x,y,z)$ at an instant of time $t$. Subsequently, a new configuration $\theta_f(x,y,z)$ is randomly chosen. The energy difference resulting from the change in configuration is computed from Equation (1) and the acceptance probability (say $P_f$) of the new configuration is determined using the Metropolis formula \cite{landau,binder}.
A random number ($r$) is chosen from a uniform distribution between 0 and 1. If the generated number $r\leq P_f$, the selected site is assigned to the new spin configuration $\theta_f (x, y, z, t)$ at the subsequent instant. In our numerical simulation, $L^3$ number of such spin updates collectively form one Monte Carlo Step per site (MCSS) which serves as the  unit of time. Additionally, we set $J=k=1$ to establish an appropriate temperature scale. Throughout the simulation, we conducted a total
of $t$ Monte Carlo steps per site (MCSS). From these $t$ steps we discarded initial $t'$ transient steps. The system is allowed
to achieve equilibrium after $t'$ MCSS. We verified that the initially discarded MCSS was sufficient to reach equilibrium results within the desired temperature range. The thermodynamic quantities are calculated by averaging over rest $t-t'$ 
MCSS, assuming the ergodicity,i.e., time average provides the ensemble average. In our numerical calculations 
we have considered $t$ ranging from 20000 to 36000 and $t'$ ranging from 10000 to 27000 depending on the values of system size ($L$) and
the temperature ($T$). The system suffers from critical slowing down near the critical temperature and requires huge time to
relax towards the equilibrium. The equilibrium values of macroscopic thermodynamic quantities are calculated over many (ranging from 50 to 300) samples. By cooling (with small steps of temperature) the system from a high-temperature paramagnetic phase, we obtained the quantities as a function of temperature. The following quantities are calculated:

The instantaneous components of magnetisations at any i-th lattice site are 

\begin{equation}
M_{x}= {1 \over {L^3}} \sum_i S^x_i 
= {1 \over {L^3}} \sum_i {\rm cos} (\theta_i)
\end{equation}

\noindent and 

\begin{equation}
M_{y}= {1 \over {L^3}} \sum_i S^y_i
= {1 \over {L^3}} \sum_i {\rm sin}(\theta_i).
\end{equation}

The instantaneous equilibrium magnetisation is measured as 

\begin{equation}
M=\sqrt{M_{x}^2+M_{y}^2}. 
\end{equation}

The average (over time) magnetisation $m=
<M>$.
The susceptibility is determined by 

\begin{equation}
\chi={L^3\over{kT}}(<M^2>-<M>^2). 
\end{equation}

The specific heat is measured as

\begin{equation}
C_v={L^3\over{kT^2}}(<E^2>-<E>^2),
\end{equation}

\noindent where $E$ is the energy density (energy per lattice site 
has been calculated from equation-1).
The symbol $<..>$, represents the time averaging (within the length of simulation),
which is approximately  equal to the ensemble averaging in the ergodic limit. All the measured quantities are calculated by further averaging over many (ranging from 50 to 300) random realizations of applied random fields.

\vskip 1.0 cm

\noindent {\bf {IV.  Results:}}

We commence by presenting the results corresponding to $h=0$. The thermodynamic phase transition in ferromagnetic 
system is generally studied by the temperature dependence of magnetisation $m$. The magnetisation ($m$) vanishes at the critical point. The susceptibility ($\chi$) and the specific-heat ($C_v$) show divergence at the critical point. However, in a
finite sized system the transition temperature or the pseudocritical temperature is determined from the maxima of the
susceptibility and the specific-heat. Our primary goal is to find the pseudocritical temperature and its dependence
on the anisotropy.

\begin{figure}[h]
\begin{center}

\resizebox{15cm}{!}{\includegraphics[angle=0]{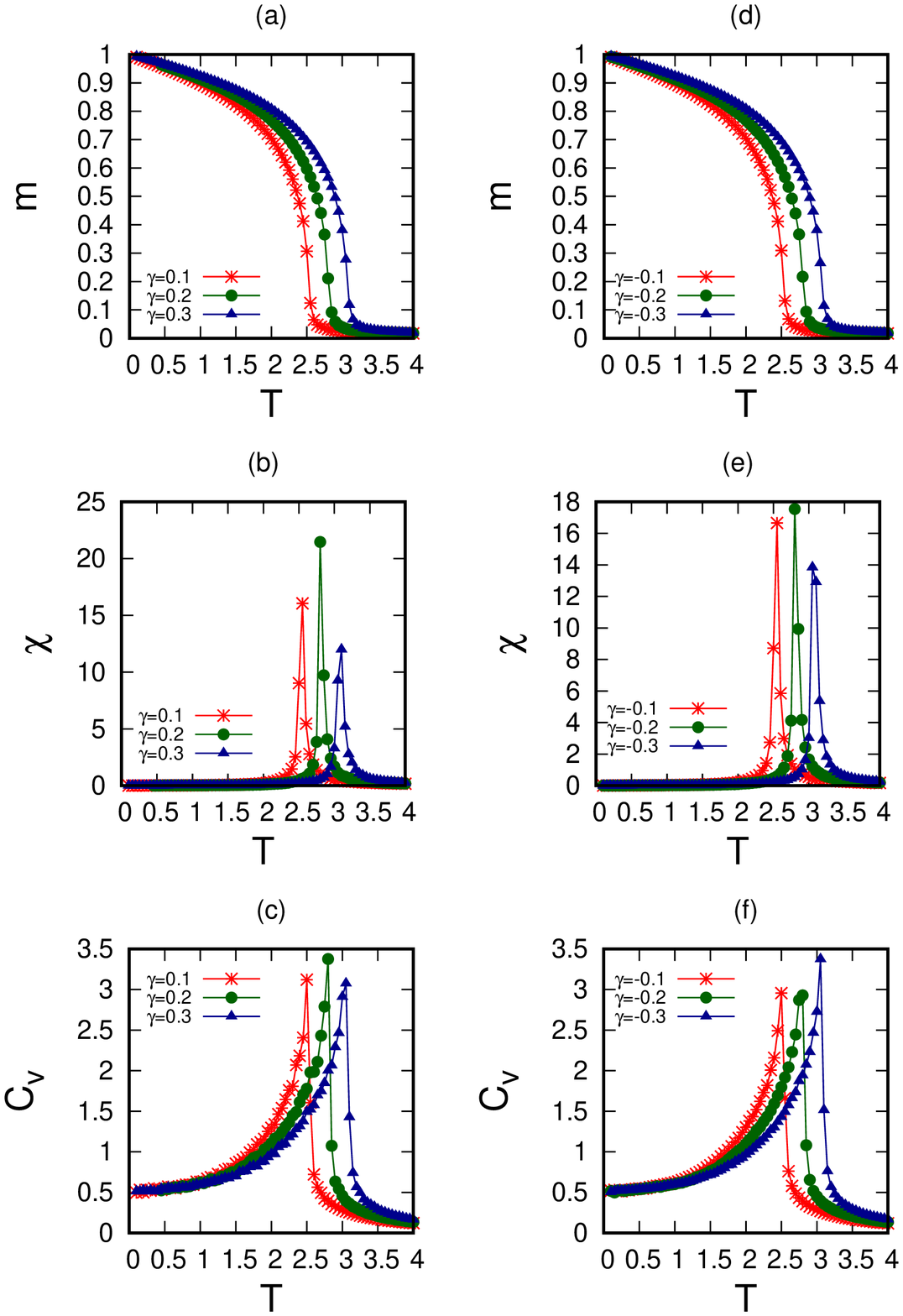}}

\caption{ The magnetisation ($m$), susceptibility ($\chi$ ) and the specific heat ($C_{v}$) are plotted as function of temperature ($T$), for three positive  as well as three negative strengths of anisotropy($\gamma$). The left panel
((a) magnetisation, (b)susceptibility and (c)specific heat) shows the results of positive anisotropies ($\gamma=0.1,0.2$ and 0.3). The right panel ((d)magnetisation, (e)susceptibility
and (f)specific heat) shows the results for negative anisotropies ($\gamma=-0.1,-0.2$ and -0.3). In both (positive and
negative anisotropy) cases, the transition (peak position of susceptibility or specific heat) occurs at higher temperature for stronger magnitude of anisotropy.}

\label{h=0.eps}
\end{center}
\end{figure}

Fig-\ref{h=0.eps} shows the thermal variations of magnetisation, susceptibility and the specific heat for two types of constant
(uniform over the space) anisotropy, namely, positive and negative. For both cases, as we decrease the temperature magnetisation grows, and susceptibility (and the specific heat also) exhibit  pronounced peaks at distinct temperatures. The system becomes ferromagnetically ordered at low temperatures.  Interestingly, for both positive and negative anisotropy,  the susceptibility peak shifts towards high temperature with increasing value of $\gamma$. The pseudocritical temperature increases with anisotropy potency. We have already reported (for positive constant anisotropy) recently\cite{olivia}. It
may be worth mentioning here that the reduction of the critical temperature was found\cite{olivia} in the case of distributed anisotropy.
However, in this article, we are considering only the constant (over the lattice) anisotropy. 

\begin{figure}[h!]

\includegraphics[angle=270,width=0.9\textwidth]{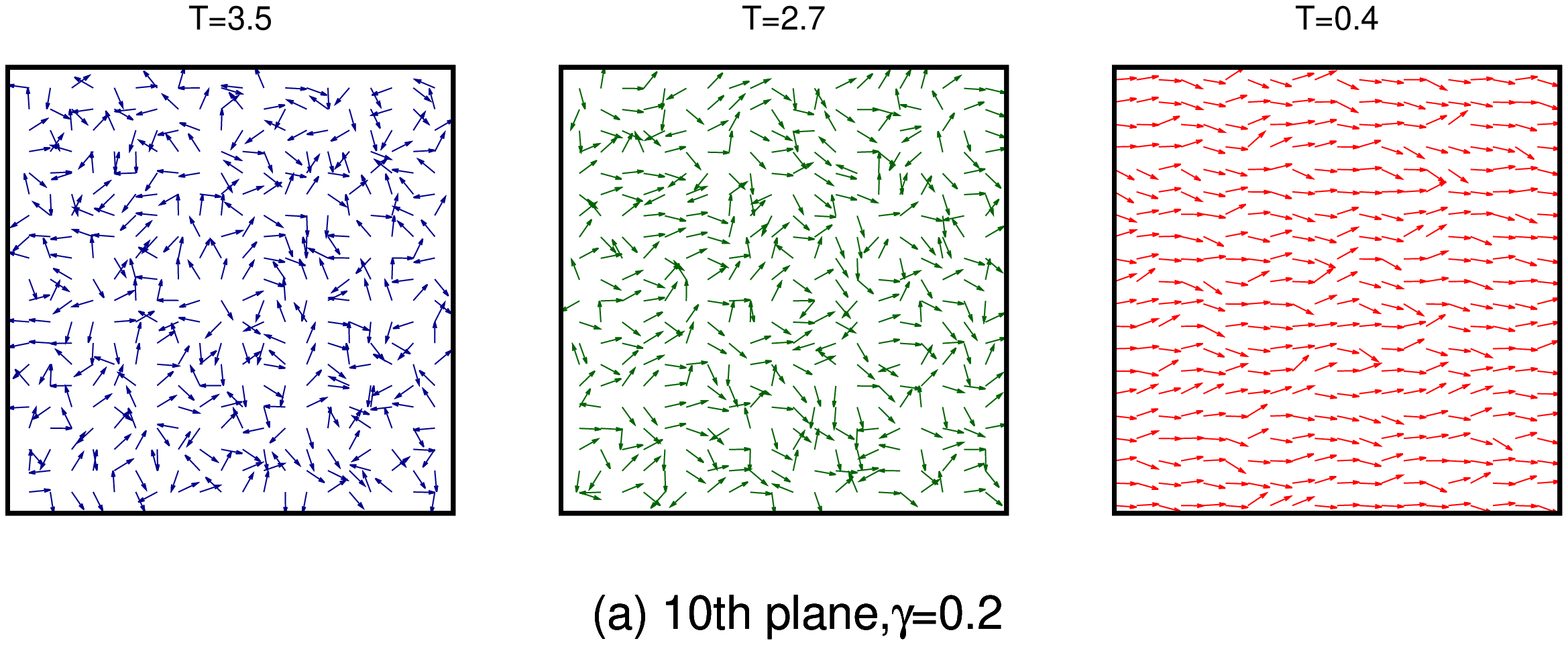}
\includegraphics[angle=270,width=0.9\textwidth]{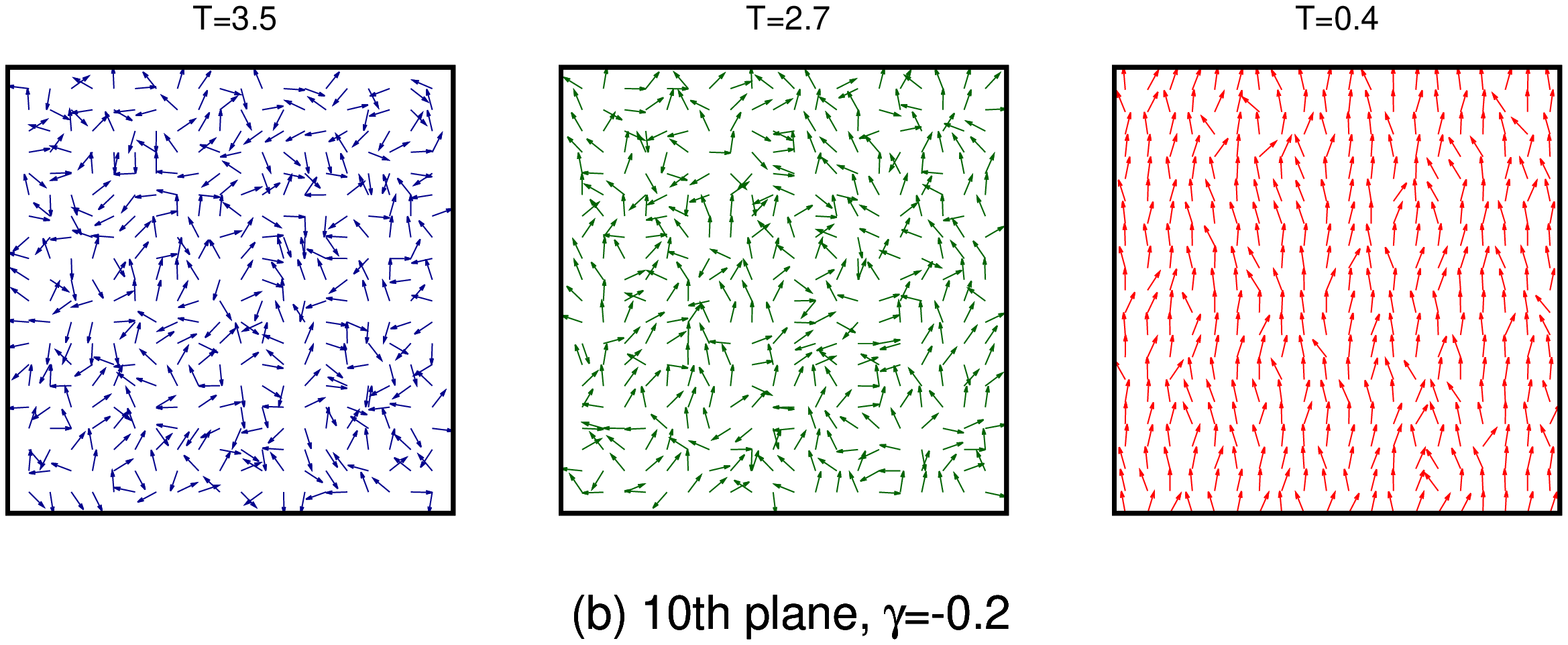}
		
	\caption{Evolution of spin configuration in XY plane at 10th plane with decreasing temperature for (a)upper pannel: positive anisotropy $\gamma=0.2$ and (b) lower pannel: negative anisotropy $\gamma=-0.2$ in absence of random field. The three temperatures ($T$) correspond to : high  $T=3.5$, near transition $T=2.7$ and low $T=0.4$ values.}
	\label{spinstructure.eps}
\end{figure}

\begin{figure}[h]
\begin{center}

\resizebox{15cm}{!}{\includegraphics[angle=270]{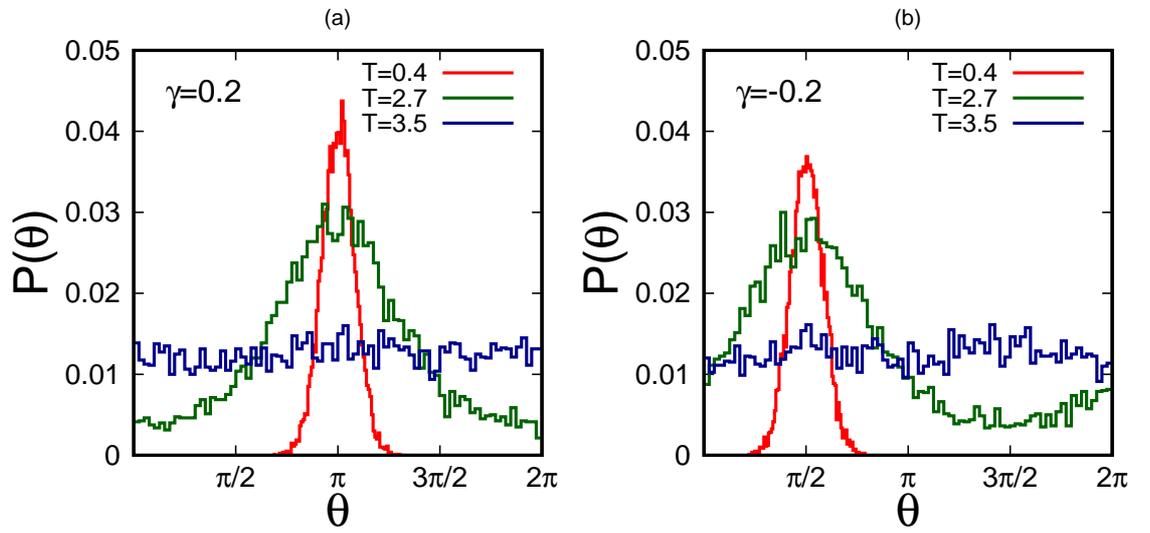}}

\caption{The normalized statistical distribution of the angular orientations of the spin vector (for a single sample) for different temperatures ($T$). (a) 
For positive anisotropy ($\gamma=0.2$), the most probable low temperature spin configuration is axially dominated along the 
x-direction ($\theta=\pi$, the configuration of $\theta=0$ is also equally probable as shown in Fig-\ref{spinstructure.eps}(a) for $T=0.4$), (b) For negative anisotropy ($\gamma=-0.2$), the most probable low temperature spin configuration is axially dominated along the 
y-direction ($\theta={{\pi} \over {2}})$.}

\label{angledist.eps}
\end{center}
\end{figure}

How can one get an idea about the orientation of spins in ordered phase depending on the sign (positive or negative) of constant anisotropy ?
Fig-\ref{spinstructure.eps} demonstrates such a spin configurations for different temperatures. The upper panel (Fig-\ref{spinstructure.eps}(a) represents the results for positive anisotropy ($\gamma=0.2$). Here, the
horizontal axis (X-axis) dominates the spin ordering at low temperature.
On the other hand, for negative anisotropy ($\gamma=-0.2$, shown in lower panel, Fig-\ref{spinstructure.eps}(b)) spins prefer to be ordered along vertical direction. This spin ordering can also be visualized statistically.
 We have studied the statistical distribution of the angular orientation of the classical spin vectors. The
low temperature phase and its orientation can be predicted from the peak position of the statistical distribution of the
angle ($\theta$).
The normalized distribution of spin angles ($\theta$) for a single sample are shown in Fig-\ref{angledist.eps}. The distribution of $\theta$, overall spins ($L^3=8000$) are shown for three different temperatures. At low temperature, for $\gamma=0.2$, the distribution gets sharply peaked near $\theta=\pi$. The distributions of angle $\theta$ assure the influence of directional predilection governed by the anisotropy. For any other independent sample, different kinds of distribution of the angles are equally probable which results the peak near $\theta=0$. Similarly, for $\gamma$=-0.2  the distribution of angles gets peak near $\theta=\pi/2$ as Y direction is preferred. $\theta=3\pi/2$ is also equally probable (for another independent sample).
Till now we discuss how anisotropy in the XY ferromagnet influences directional preferences on the behaviour and properties of the system such as phase transition, and the critical phenomena. Understanding and characterizing the effect of the anisotropy in the XY model is crucial for studying the system's behaviour and its response to the field. It provides insight into the interplay between anisotropy  and random fields, leading to a comprehensive understanding of phase transition. 

\vskip 0.5cm
\noindent\textbf{\textit{A. Random Field with full circular Symmetry:}}

\begin{figure}[h]
\begin{center}

\resizebox{8cm}{!}{\includegraphics[angle=270]{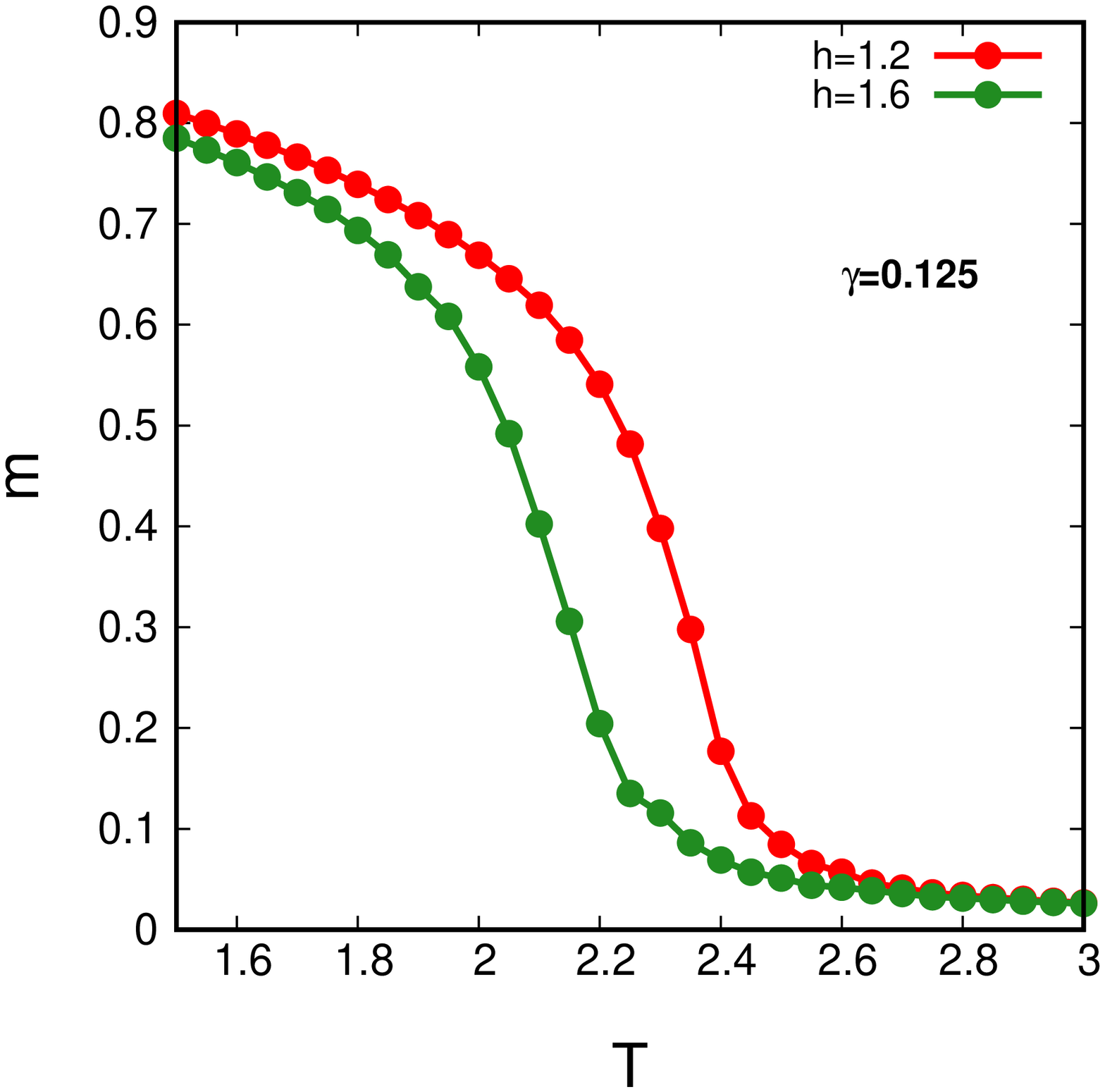}}
(a)
\resizebox{8cm}{!}{\includegraphics[angle=270]{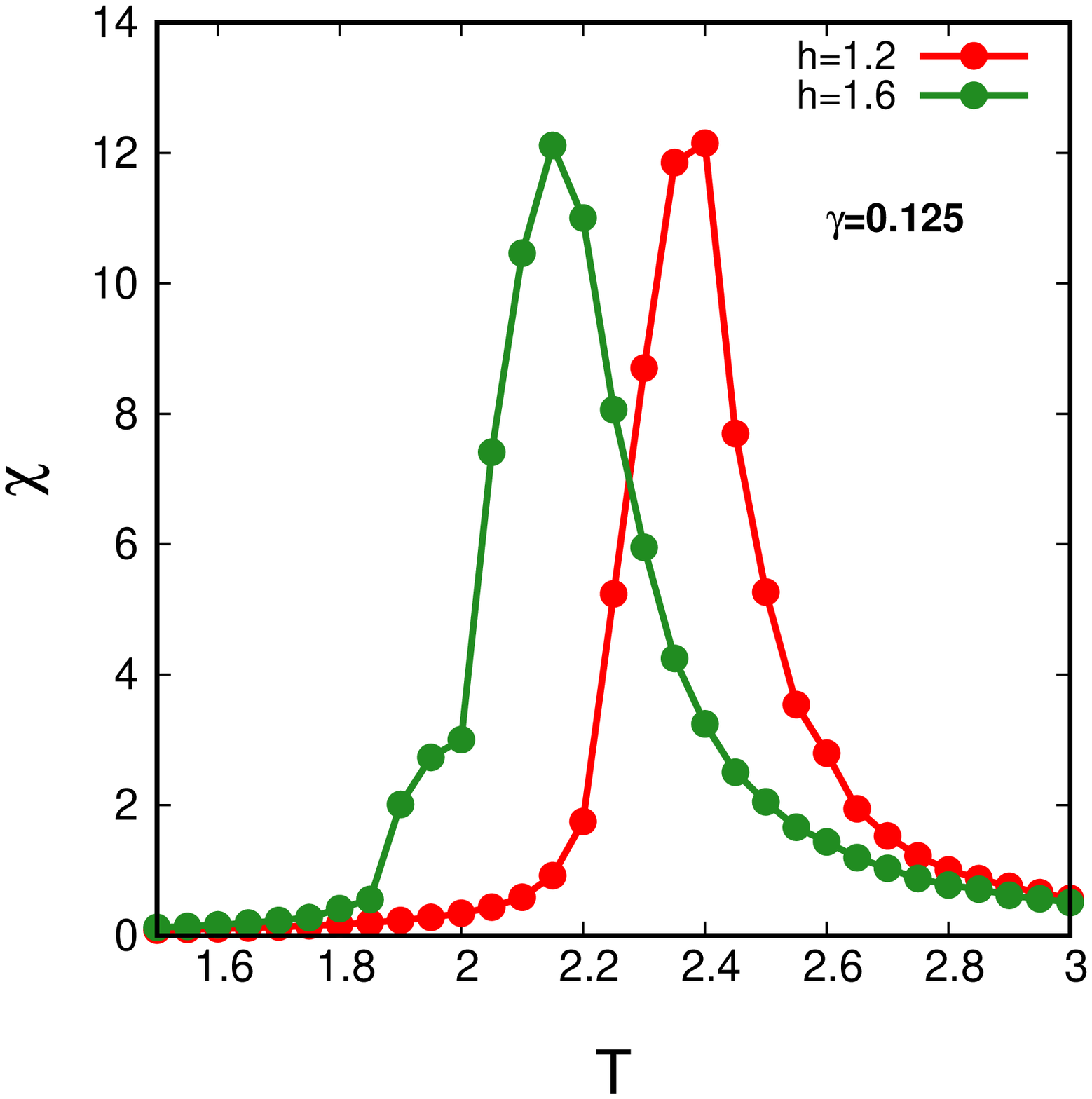}}
(b)
\resizebox{8cm}{!}{\includegraphics[angle=270]{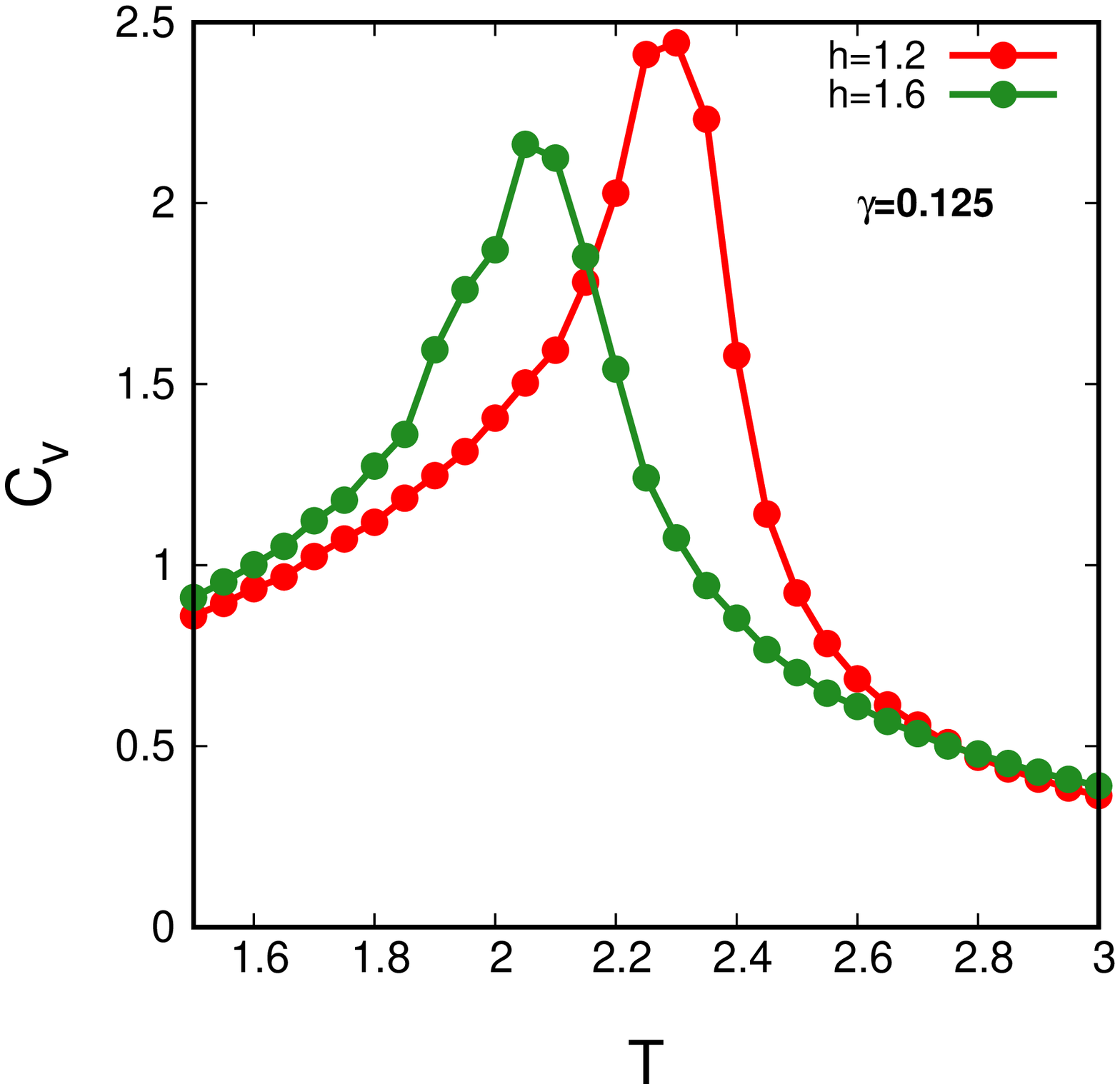}}
(c)

\caption{The (a) magnetization ($m$), (b) the susceptibility ($\chi$) and
(c) the specific-heat ($C_v$) (c) are plotted against the temperature ($T$) for two different strengths of the uniform random fields,
$h=1.2$ and $h=1.6$ with ($\gamma=0.125$).}

\label{m-g0.125.eps}
\end{center}
\end{figure}

How does the random field affect the critical behaviour of the three dimensional anisotropic XY ferromagnet ? To address this question
we consider a random field of fixed magnitude ($h$) but the direction (angle $\phi_i$) is distributed uniformly between 0 to $2\pi$. Let us call this A-type random field. For this kind of randomness of the field the average field $<\vec h_i>=0$. 
The random disorder (here the random field) having a null gross effect ($<\vec h_i>=0$) may strongly govern the critical behaviour
of the anisotropic XY ferromagnet.
We are interested to study the effect of such kind of random field
 (random disorder) on the critical behaviour of anisotropic XY ferromagnet in three dimensions. We have studied the thermal variation of magnetisation, susceptibility and specific heat for fixed anisotropy ($\gamma$) values in Fig-\ref{m-g0.125.eps}. As we cool the system from high temperature paramagnetic phase, magnetization grows. At low temperature regime spins get ordered. The thermal variation of magnetic susceptibility is shown in Fig-\ref{m-g0.125.eps}(b). The pseudo-critical temperatures (critical temperature for finite sized system) is obtained from the positions of peaks of the susceptibility. 
We observed that for a particular anisotropy (say $\gamma=0.125$), the pseudo-critical temperature decreased as the  strength of the random field increased. The variations of magnetisation, susceptibility and the specific heat with temperature (at $\gamma=0.125$) for two different values ($h=1.2,h=1.6$) of the strengths of the random field, are shown in Fig-\ref{m-g0.125.eps}. The stronger random fields are found to be responsible for lower pseudocritical
 temperature of anisotropic XY ferromagnet in three dimensions. Here, the random field is acting like the quenched random disorder and hence  reduces the pseudocritical temperature ($T_c^{*}$).

\begin{figure}[h]
\begin{center}

\resizebox{15cm}{!}{\includegraphics[angle=270]{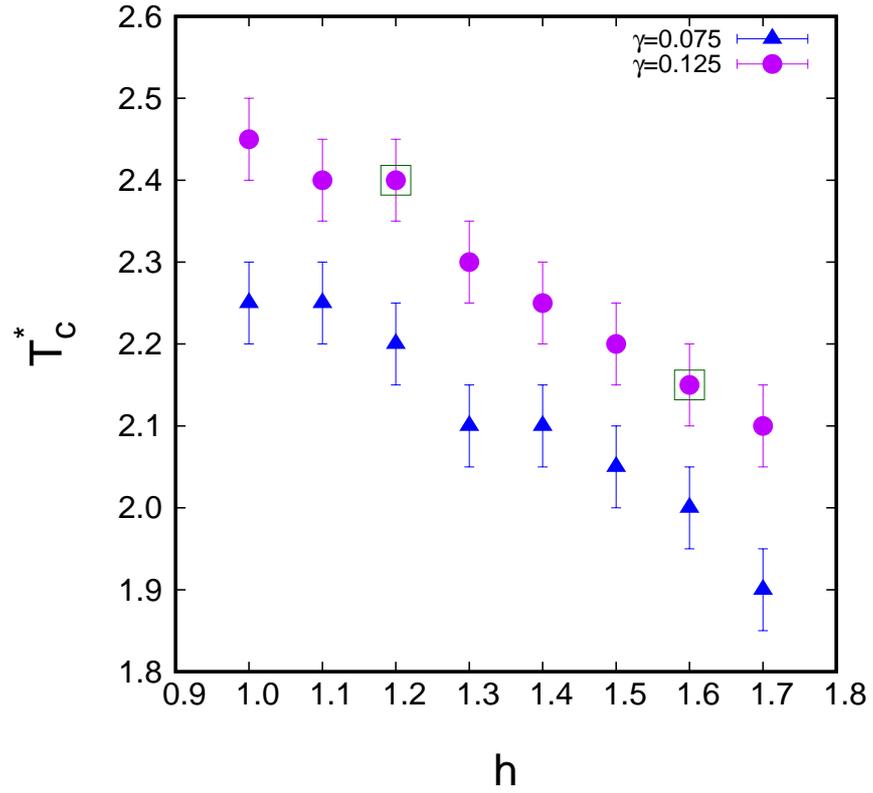}}

\caption{ The pseudo-critical temperature ($T^{*}_{c}$) (obtained from the peak position of the susceptibility) plotted against the strength ($h$) of random field (with full
circular symmetry ) for different strengths of the anisotropy ($\gamma$). The temperature dependences of the magnetisation ($m$) and the susceptibility ($\chi$)
are shown in Fig-\ref{m-g0.125.eps} at the points marked by boxes. The errorbars are the maximum range arising from the stepsize
of cooling the system.}

\label {h-Tc.eps}
\end{center}
\end{figure}

Therefore, the constant anisotropy increases critical temperature by imposing directional preference on the spins while the random field decreases critical temperature by introducing a disorder into the system.
The reduction of the pseudocritical temperature is depicted in Fig-\ref{h-Tc.eps} for two different values of the strengths of the anisotropy.
 So, it offers a competition between constant anisotropy and the random field in the context of having the critical temperatures. {\it Can the random field nullify the effect of anisotropy ? Precisely,  
is it possible to have the critical temperature for three dimensional isotropic XY ferromagnet by tuning the values of the strengths of the random field and that of the constant anisotropy ?} The strength (depending on the strength of anisotropy) of the random field required to get the value of critical temperature for three dimensional {\it isotropic} XY ferromagnet may be called the {\it compensating field} ($h_c$).

\begin{figure}[h]
\begin{center}

\resizebox{15cm}{!}{\includegraphics[angle=270]{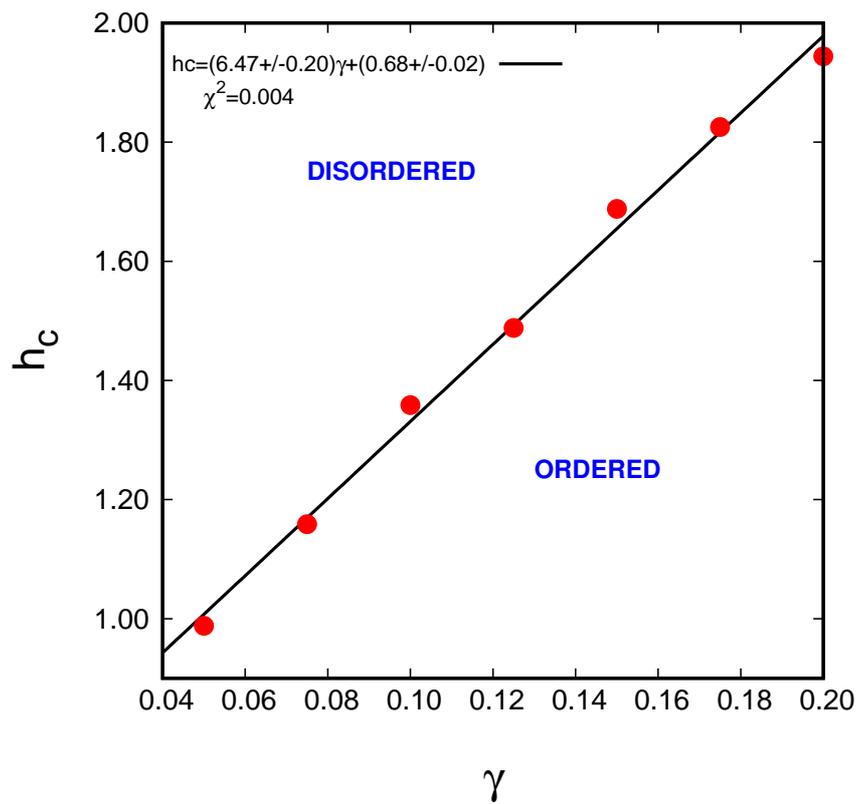}}

\caption{The compensating field ($h_c$) is plotted as function of the strength of anisotropy ($\gamma$). The
compensating field has been calculated from the linear interpolation around the true critical temperature
($T_c=2.206$)\cite{campostrini,hasenbusch}. The data points are fitted with a straight line. The region of
ordered and disordered phases are separated by this line.}

\label {compfld.eps}
\end{center}
\end{figure}

We have calculated the pseudo-critical temperature for some values of random field in presence of constant anisotropy. By interpolating them we get the values of compensating field ($h_c)$ required to get the critical temperature at $T_c=2.206$ reported\cite{campostrini,hasenbusch} for the Monte Carlo estimate of the  three dimensional isotropic XY ferromagnet. We have plotted the pairs of ($\gamma$, $h_c$) in Fig-\ref{compfld.eps}. and fitted it to straight line . Below this line we get a ferromagnetic ordered phase and above corresponds to disordered phase.

\begin{figure}[h]
\begin{center}

\resizebox{15cm}{!}{\includegraphics[angle=270]{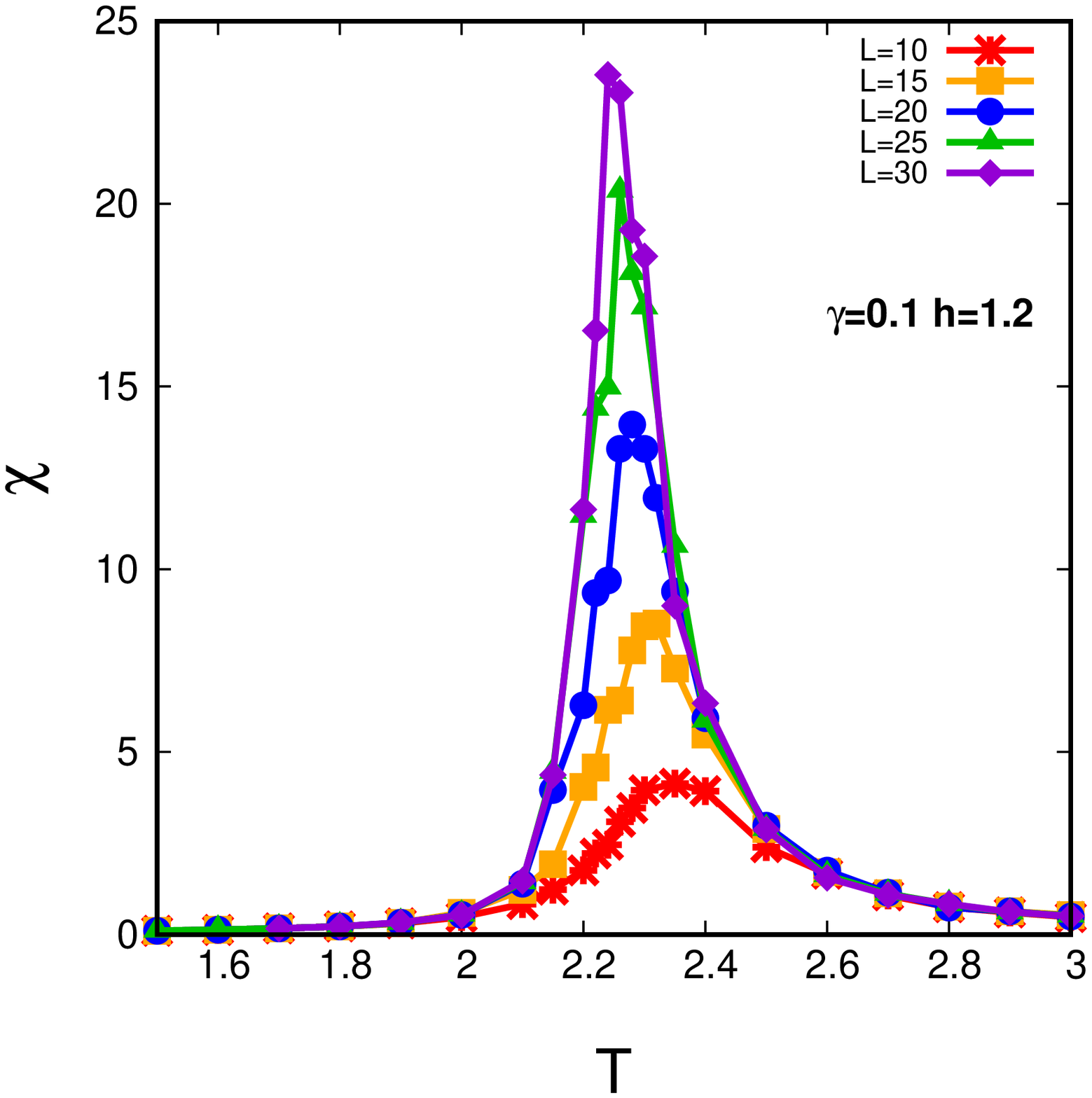}}

\caption{The susceptibility ($\chi$) plotted as function of the temperature ($T$) for five
different system sizes ($L=10,15,20,25$ and 30). Here, magnitude of the random
field, $h=1.2$ and the strength of the anisotropy, $\gamma=0.1$. The random field (A-type) has the direction
chosen randomly within angle 0 and $2\pi$.
}

\label {chi-T-circular.eps}
\end{center}
\end{figure}

\begin{figure}[h]
\begin{center}

\resizebox{15cm}{!}{\includegraphics[angle=270]{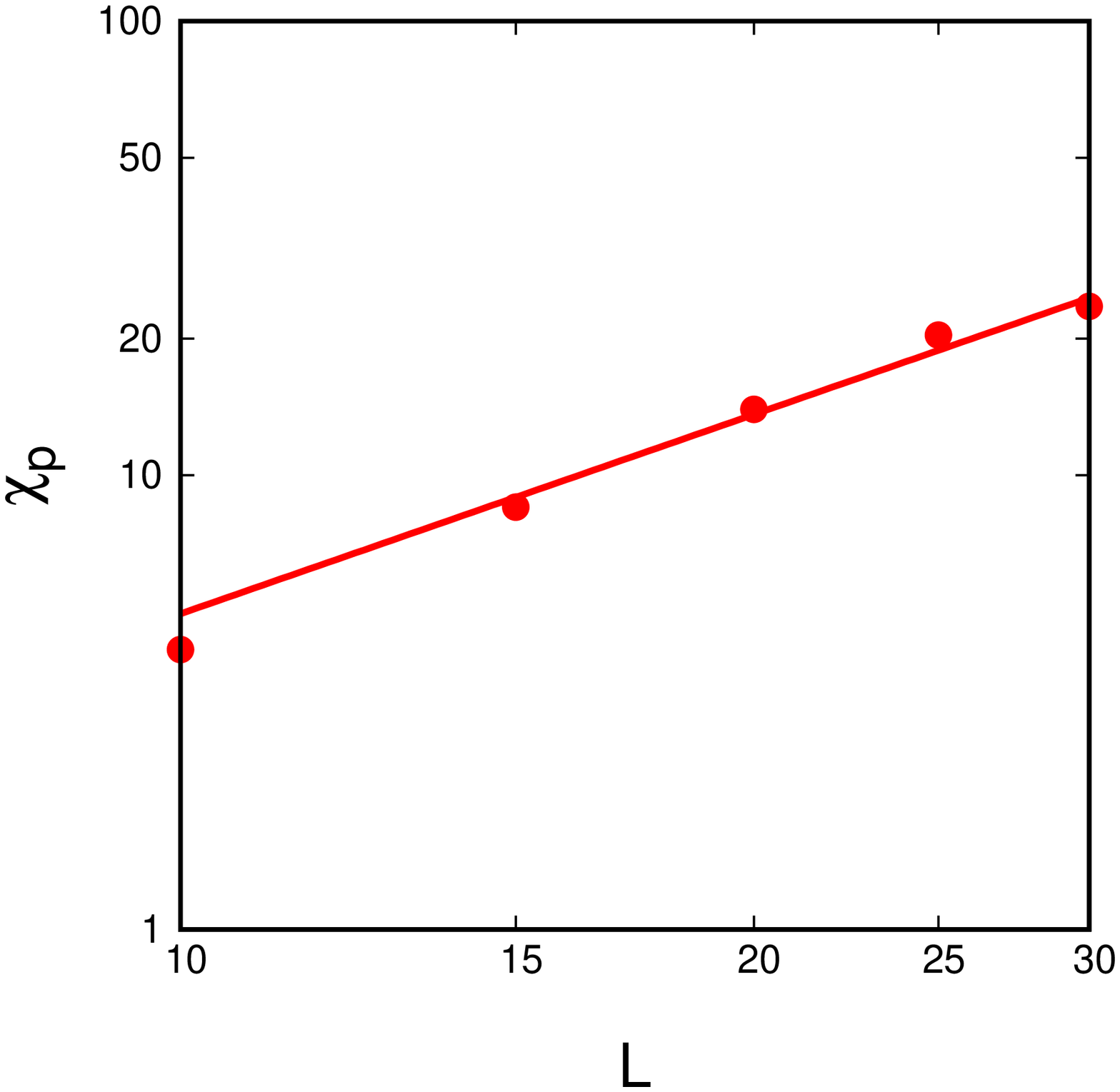}}

\caption{The logarithm of the heights of the peaks of the susceptibilities ($\chi_p$) are plotted against
the logarithm of system sizes $L$, for $h=1.2$ and $\gamma=0.1$. The data points are fitted with a straight
line assuming the scaling law $\chi_p \sim L^{{\gamma'} \over {\nu}}$. Here, we have estimated (from the linear
best fit) ${{\gamma'} \over {\nu}}=1.46\pm0.14$. Here, the A-type random field is used.}

\label {chi-peak-L.eps}
\end{center}
\end{figure}

\begin{figure}[h]
\begin{center}

\resizebox{15cm}{!}{\includegraphics[angle=270]{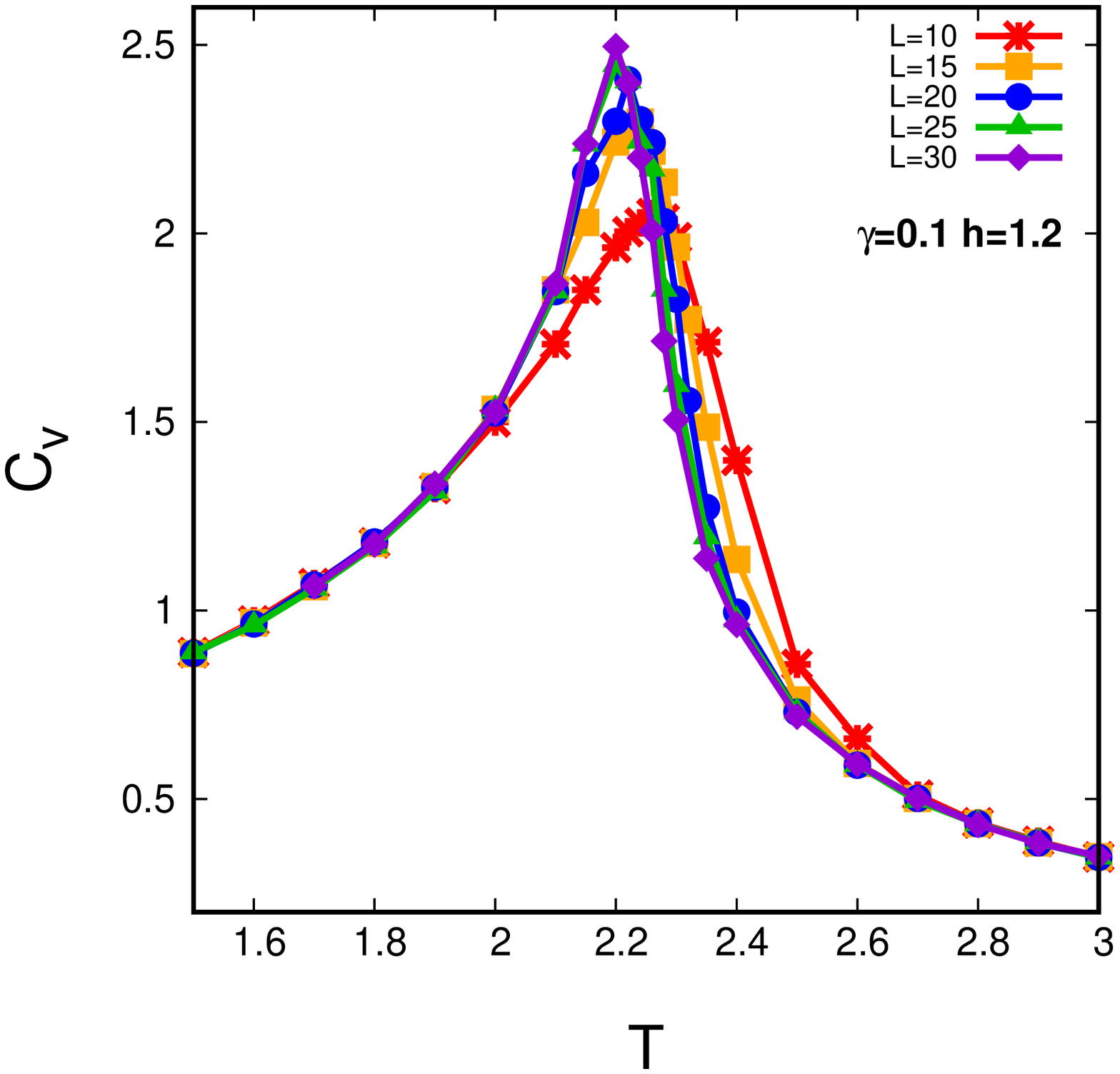}}

\caption{The specific heat ($C_v$) plotted as function of the temperature ($T$) for five
different system sizes ($L=10,15,20,25$ and 30). Here, magnitude of the random
field, $h=1.2$ and the strength of the anisotropy, $\gamma=0.1$. The random field (A-type) has the direction
chosen randomly within angle 0 and $2\pi$.
}

\label {cv-T-circular.eps} 
\end{center}
\end{figure}

\begin{figure}[h]
\begin{center}

\resizebox{15cm}{!}{\includegraphics[angle=270]{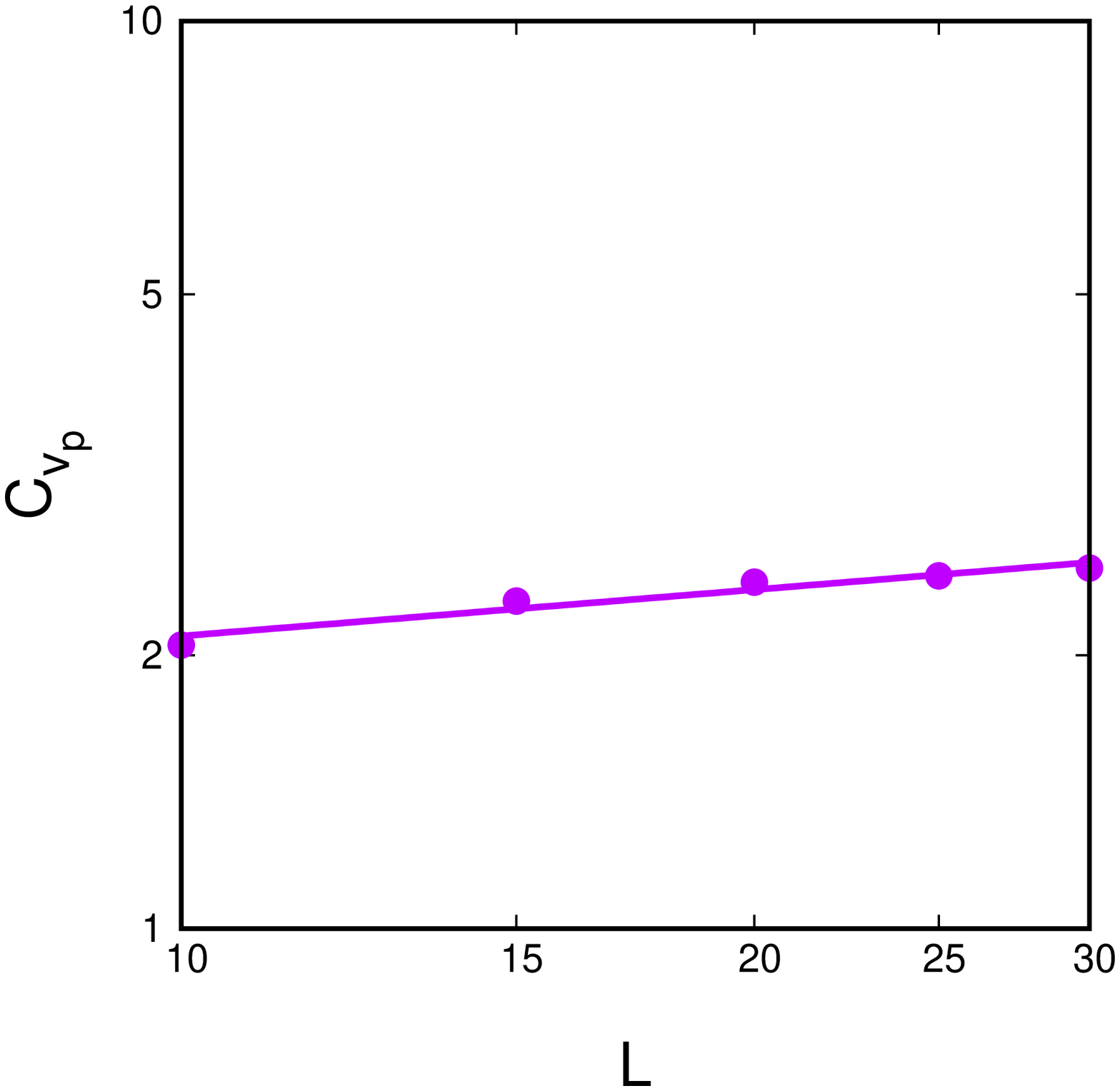}}

\caption{The logarithm of the heights of the peaks of the specific-heat ($C_{vp}$) are plotted against
the logarithm of system sizes $L$, for $h=1.2$ and $\gamma=0.1$. The data points are fitted with a straight
line assuming the scaling law $\chi_p \sim L^{{\alpha} \over {\nu}}$. Here, we have estimated (from the linear
best fit) ${{\alpha} \over {\nu}}=0.17\pm0.02$. Here, the A-type random field is used }

\label {cv-peak-L.eps}
\end{center}
\end{figure}

Any critical behaviour or thermodynamic phase transition should be formalized by the finite size study. Generally, it is customary to check whether
the effect of the critical correlations diverges at the critical point or not in the thermodynamics limit ($L \to \infty$). Here
also we have studied the thermal variations of susceptibility ($\chi$) and the specific heat ($C_v$) for different system sizes
($L=10,15,20,25$ and 30) and at fixed  $h=1.2$ and $\gamma=0.1$. Fig-\ref{chi-T-circular.eps} shows the temperature
dependence of susceptibility ($\chi$) for different system sizes. It is clear from the figure that the peak height of the suceptibility increases as the system size ($L$) increases. This is the signature of the growth of the critical correlation.
The susceptibility shows the tendency of divergence at the transition point. The height of the peak 
of the susceptibility ($\chi_p$) is plotted with the system size ($L$) and shown in Fig-\ref{chi-peak-L.eps}. It may be noted here that to identify the peak precisely, the step-size of the temperature has been reduced to
$\Delta T=0.02$, in the vicinity of the peak position. Assuming a
scaling form $\chi_p \sim L^{{\gamma'} \over {\nu}}$, the data are fitted. This best fit (with $\chi^2=4.39$, DOF=3) estimates
${{\gamma'} \over {\nu}}=1.46\pm0.14$. We have studied the thermal variation of the specific-heat ($C_v$)
and shown in Fig-\ref{cv-T-circular.eps}. Here also, the height of the peak of the specific heat
($C_{vp}$) was found to increase as the
system size ($L$) is increased. Assuming the scaling form $C_{vp} \sim L^{{\alpha} \over {\nu}}$, the data are shown in 
Fig-\ref{cv-peak-L.eps}. The best fit estimated the exponent ${{\alpha} \over {\nu}}=0.17\pm0.02$ (with $\chi^2=0.007$, DOF=3).
It may be noted here that such small value of the exponent (${{\alpha} \over {\nu}}$ ) was also estimated
(${{\alpha} \over {\nu}}=0.02$) \cite{campostrini}
in three dimensional isotropic XY ferromagnet by Monte Carlo simulation.

\vskip 0.5cm

\noindent\textbf{\textit{B. Random Field within specified angular window:}}

\begin{figure}[h]
\begin{center}

\resizebox{10cm}{!}{\includegraphics[angle=0]{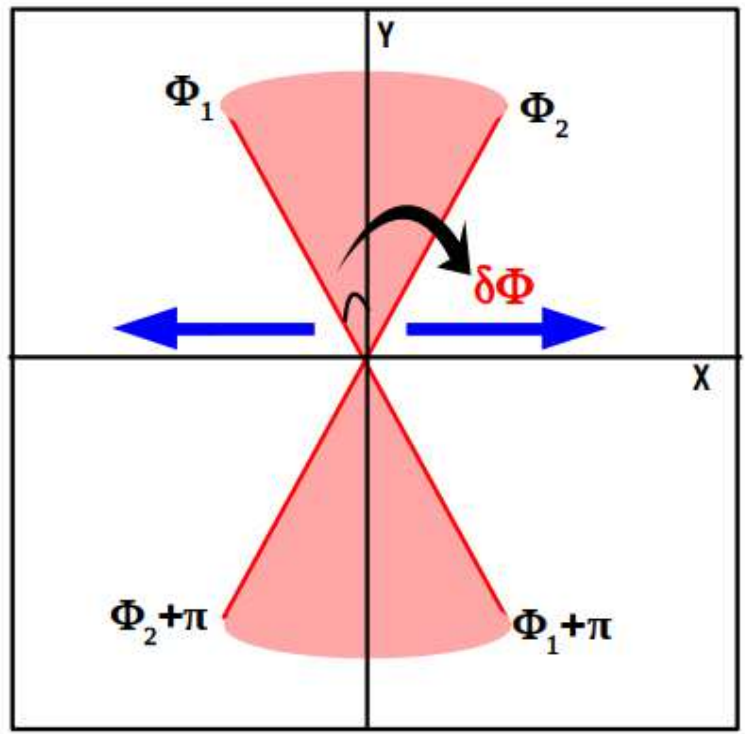}}

\caption{A schematic diagram to demonstrate the angular region of the direction of the random field. Here, the
random field cannot take any direction between 0 and $2\pi$. The blue arrows represent the dominance 
of the directions of preferred ordering of spin vectors due to positive anisotropy.}

\label {ang-window.eps}
\end{center}
\end{figure}

From the discussion above, key findings emerge that the directional predilection imposed by constant anisotropy can be crank down by applying a uniform random field with circular symmetry (0 to $2\pi$). Now let us see what happens to 
the critical behaviours of the anisotropic XY ferromagnet (in three dimensions) if the random field is circumscribed about an angular window. Let us call it B-type random field. Here also, the average field $<\vec h_i>=0$. We have demonstrated the range of allowed directions of such kind of random field in
Fig-\ref{ang-window.eps}. It may be noted here that B-type random field maps onto A-type random field in the limit
$\delta\phi \to 0.5\pi$. In accordance with previous studies, the constant positive $\gamma$ governs the spins to orient in X direction, we apply the random field in transverse conic as shown in Figure. The width of the angular window is defined as $2\delta\Phi=\Phi_{1}-\Phi_{2}$, $\delta\Phi$ is measured from Y axis to field direction. The random field with constant magnitude $h$ is applied in both positive and negative Y direction within the conic (coloured region). First we consider a field strength $h=1.4$ in presence of constant $\gamma=0.15$. The thermal variation of magnetization, susceptibility and specific heat for two different angular windows $\delta\Phi=0.15\pi$ and $\delta\Phi=0.40\pi$ are depicted in Figure. The pseudo-critical temperature is obtained from the  position of peak of the susceptibility. The pseudo-critical temperature is found to increase by constricting the width of angle window. The variation of pseudo-critical temperature with $\delta\Phi$ for three different $\gamma$ is shown in Figure. As we start to constrict the random field angle ($\Phi$) from circular symmetry ($4\delta\Phi=2\pi$) to the angular window , the pseudo-critical temperature begins to elevate monotonically. If we decrease anisotropy strength from $\gamma=0.15$ to $\gamma=0.05$
the line of pseudo-critical temperature  goes down as shown in Fig-\ref{ang-Tc.eps}.
One may think it in a different way; if we increase the angular window from $\delta\Phi=0.1\pi$ to $\delta\Phi=0.5\pi$ the pseudo-critical temperature decreases by approaching the circular symmetry. For $\gamma=0.1$ at $\delta\Phi=0.5\pi$ the pseudo-critical temperature $T^*_{c}=2.20$ corresponds to circular symmetry that maximizes the effect of random field. Therefore the random field restricted in an angular window enables to reduce the critical temperature of anisotropic XY ferromagnet.

\begin{figure}[h]
\begin{center}

\resizebox{15cm}{!}{\includegraphics[angle=270]{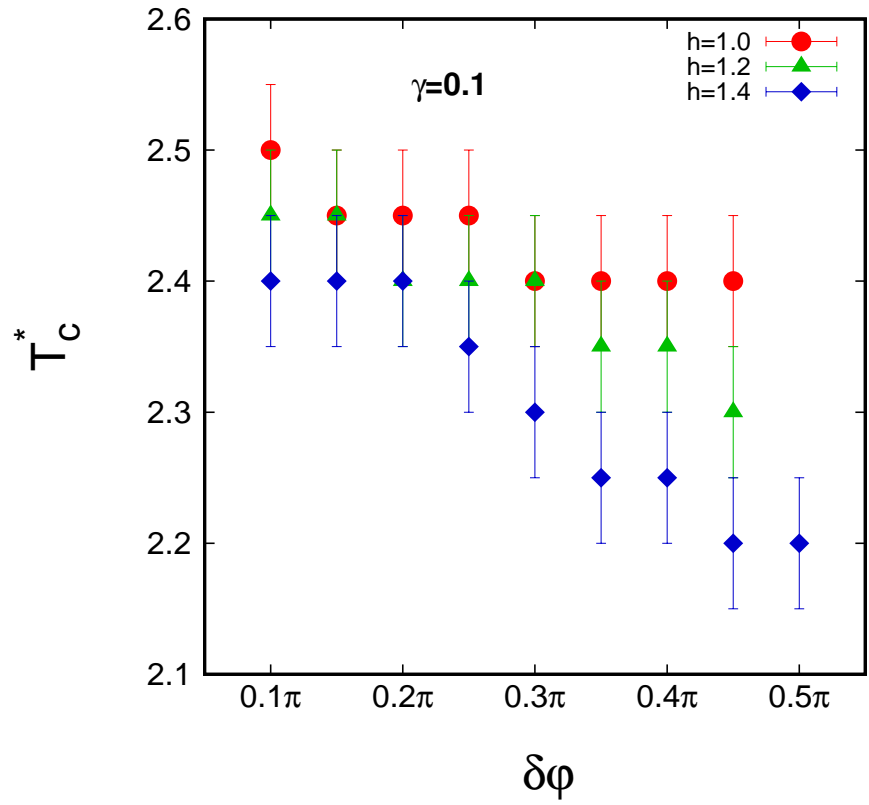}}

\caption{The variation of the pseudocritical temperature ($T_{c}^{*}$) with the angular extension ($\delta\phi$) 
for constant anisotropy $\gamma=0.1$. The three different symbols represents three different
values of the strength of the random field. The $4\delta\phi=2\pi$ restores the full circular symmetry of
the random field. The errorbars are the maximum range arises from the stepsize of cooling
the system.}

\label {ang-Tc-const-g.eps}
\end{center}
\end{figure}

For fixed strength of the anisotropy ($\gamma=0.1$), the variation of the
pseudocritical temperature ($T_c^{*}$) has been studied as function of the
angular extension ($\delta\Phi$) for three different magnitudes
($h=1.0,1.2$ and 1.4) of B-type random
fields. The results are shown in Fig-\ref{ang-Tc-const-g.eps}. For stronger
magnitude of random field, the pseudocritical temperature 
($T_c^{*}$) has been found to decrease with increasing the angular extension ($\delta\Phi$). The pseudocritical temperature for A-type random field has been restored for $\delta\Phi=0.5\pi$ (for $h=1.4$ shown by blue diamonds in the diagram). However, for weaker fields the less amount of reduction of the pseudocritical temperature with the angular extension has been noticed.

\begin{figure}[h]
\begin{center}

\resizebox{15cm}{!}{\includegraphics[angle=270]{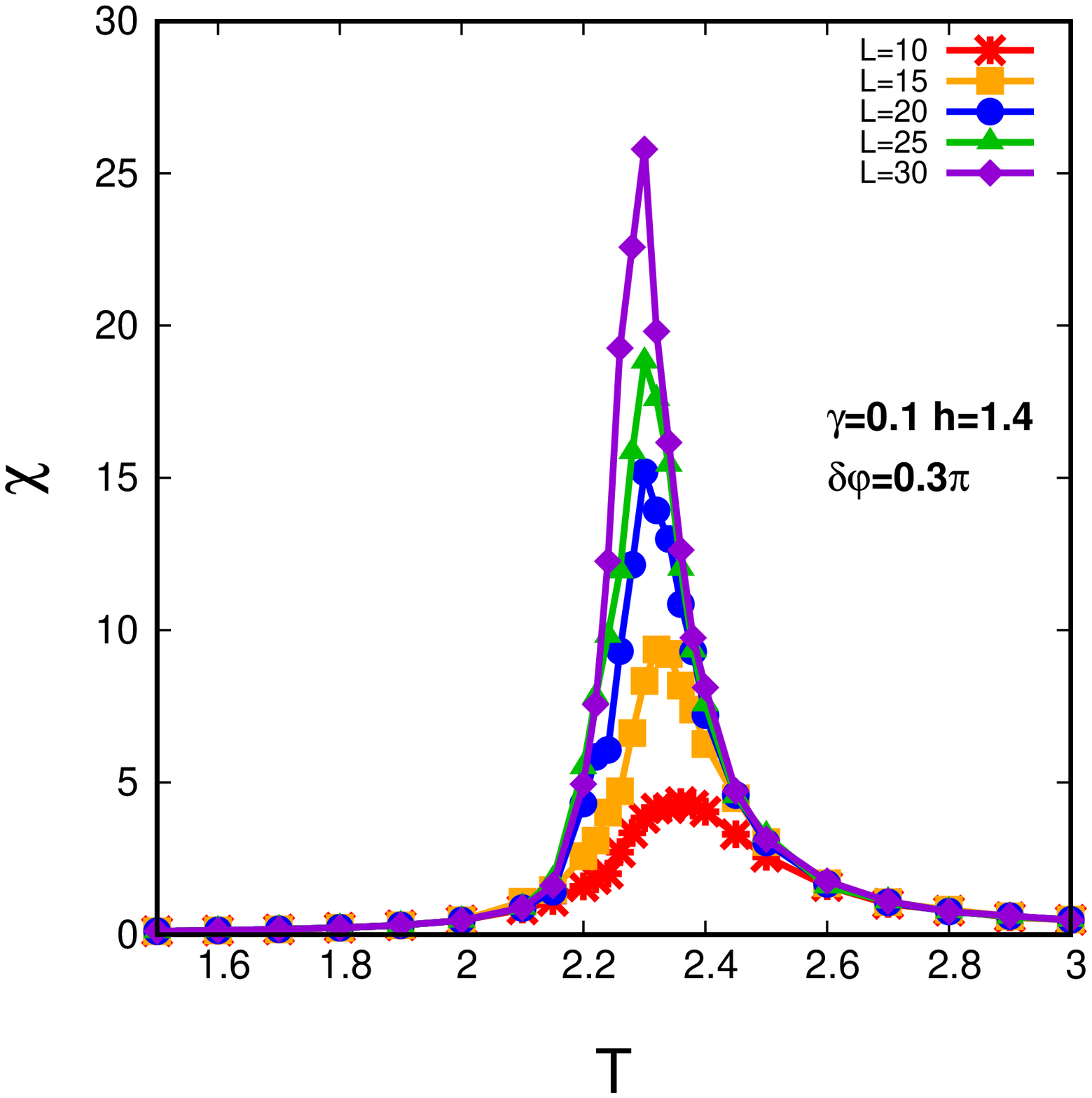}}

\caption{The susceptibility ($\chi$) plotted as function of the temperature ($T$) for five
different system sizes ($L=10,15,20,25$ and 30). Here, magnitude of the random
field, $h=1.4$, the strength of the anisotropy, $\gamma=0.1$ and $\delta\phi=0.3\pi$. 
Here, the B-type random field has been used.  }

\label {chi-T-conical.eps}
\end{center}
\end{figure}

\begin{figure}[h]
\begin{center}

\resizebox{15cm}{!}{\includegraphics[angle=270]{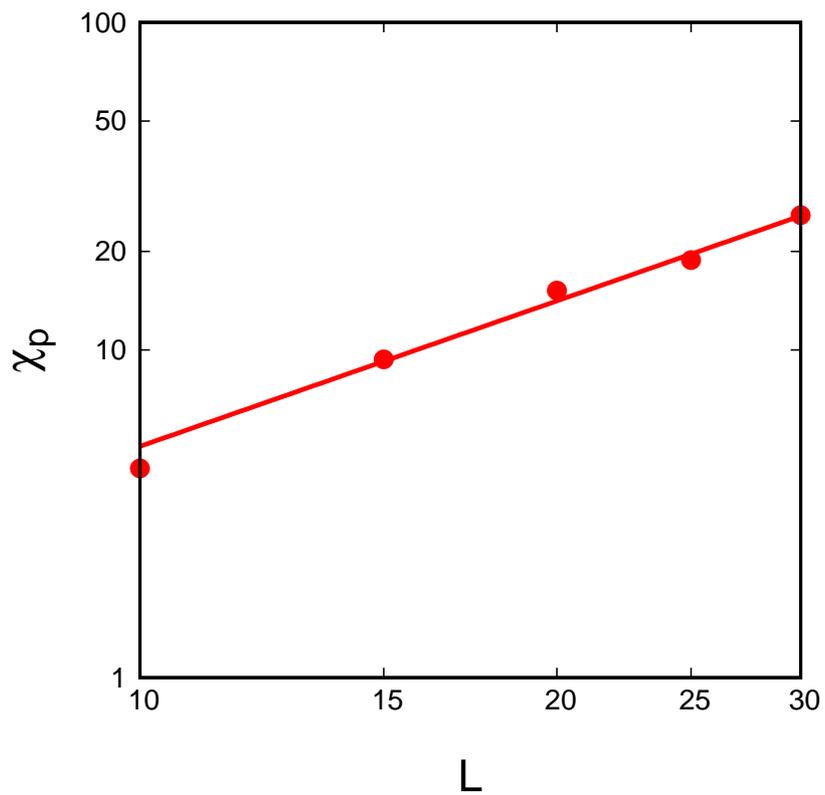}}

\caption{The logarithm of the heights of the peaks of the susceptibilities ($\chi_{p}$) are plotted against
the logarithm of system sizes L, for $h = 1.4$ ,$\gamma=0.1$ and $\delta\phi=0.3\pi$. The data points are fitted with a
straight line assuming the scaling law $\chi_{p}\sim L^{\gamma'/\nu}$ . Here, we have estimated (from the linear best fit) $\gamma' / \nu= 1.47\pm 0.10$ having $\chi^2=2.36$ and DOF=3. Here, the B-type random field is used.}

\label {chi-peak-L-conical.eps}
\end{center}
\end{figure}

\begin{figure}[h]
\begin{center}

\resizebox{15cm}{!}{\includegraphics[angle=270]{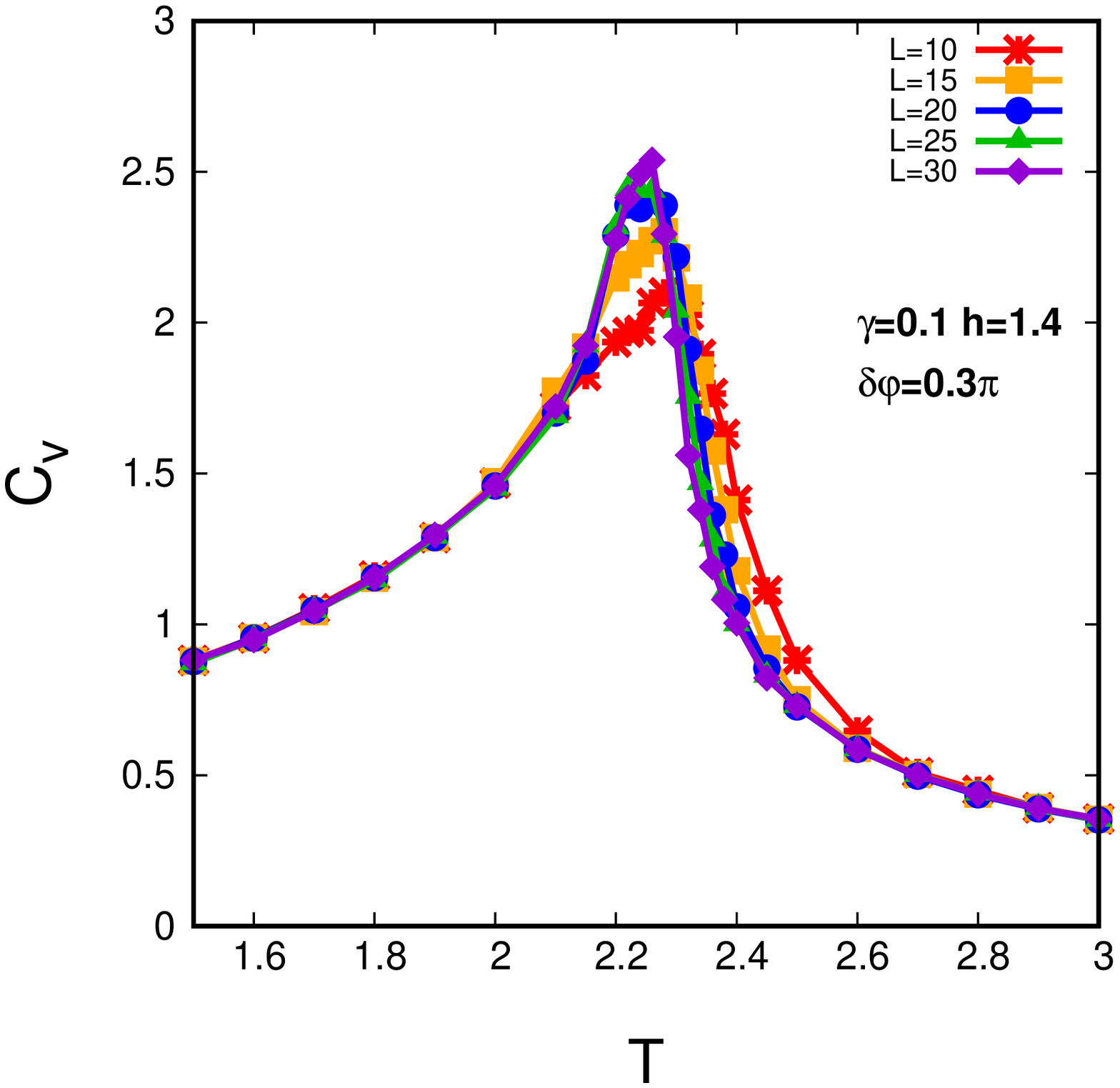}}

\caption{The specific-heat ($C_v$) plotted as function of the temperature ($T$) for five
different system sizes ($L=10,15,20,25$ and 30). Here, magnitude of the random
field, $h=1.4$, the strength of the anisotropy, $\gamma=0.1$ and $\delta\phi=0.3\pi$. 
Here, the B-type random field has been used.}

\label {cv-T-conical.eps}
\end{center}
\end{figure}

\begin{figure}[h]
\begin{center}

\resizebox{15cm}{!}{\includegraphics[angle=270]{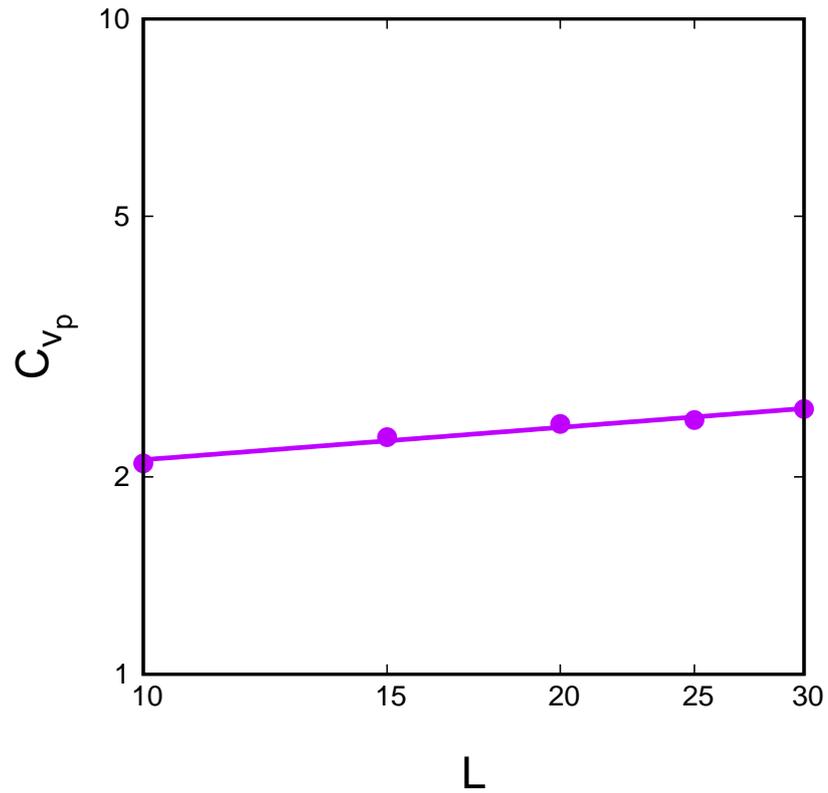}}

\caption{The logarithm of the heights of the peaks of the specific heat ($C_{vp}$) are plotted against
the logarithm of system sizes L, for $h = 1.4$ ,$\gamma=0.1$ and $\delta\phi=0.3\pi$. The data points are fitted with a
straight line assuming the scaling law $C_{vp}\sim L^{\alpha/\nu}$ . Here, we have estimated (from the linear best fit) $\alpha / \nu= 0.16\pm 0.02$
having $\chi^2=0.003$ and DOF=3. Here, the B-type random field is used.}

\label {cv-peak-L-conical.eps}
\end{center}
\end{figure}

Here
also we have studied the thermal variations of susceptibility ($\chi$) and the specific heat ($C_v$) for different system sizes
($L=10,15,20,25$ and 30) and for fixed values of $h=1.4$, $\gamma=0.1$ and $\delta\Phi=0.3\pi$. Fig-\ref{chi-T-conical.eps} shows the temperature
dependence of susceptibility ($\chi$) for different system sizes. It is clear from the figure that the peak height of the susceptibility increases as the system size ($L$) increases. This is the signature of the growth of the critical correlation.
The susceptibility shows the tendency of divergence at the transition point. The height of the peak 
of the susceptibility ($\chi_p$) is plotted with the system size ($L$) and shown in Fig-\ref{chi-peak-L-conical.eps}. Assuming a
scaling form $\chi_p \sim L^{{\gamma'} \over {\nu}}$, the data are fitted. This best fit (with $\chi^2=2.30$, DOF=3) estimates
${{\gamma'} \over {\nu}}=1.47\pm0.10$. We have studied the thermal variation of the specific-heat ($C_v$)
and shown in Fig-\ref{cv-T-conical.eps}. Here also, the height of the peak of the specific heat
($C_{vp}$) was found to increase as the
system size ($L$) is increased. Assuming the scaling form $C_{vp} \sim L^{{\alpha} \over {\nu}}$, the data are shown in 
Fig-\ref{cv-peak-L-conical.eps}. The best fit estimated the exponent ${{\alpha} \over {\nu}}=0.16\pm0.01$ (with $\chi^2=0.007$, DOF=3).
It may be noted here that such small value of the exponent (${{\alpha} \over {\nu}}$ ) was also estimated
(${{\alpha} \over {\nu}}=0.02$, where $\alpha=0.0146$ and $\nu=0.6715$) \cite{campostrini}
in three dimensional isotropic XY ferromagnet by Monte Carlo simulation.


\vskip 0.8cm
\noindent {\bf V. Summary:}

We have studied the critical behaviours of anisotropic XY ferromagnet in three dimensions by Monte Carlo simulation using Metropolis algorithm. The critical behaviours of anisotropic XY 
ferromagnet has also been studied recently \cite{olivia}. The
anisotropy (breaks the SO(2) symmetry) causes the order-disorder phase transition at higher temperatures. Whereas, the randomly distributed anisotropy causes the reduction of critical temperature\cite{olivia}. What would be the effects of random magnetic fields on the critical behaviours of  the anisotropic XY ferromagnets? The three dimensional random field isotropic XY model does not
show any ferro-para phase transition. The minimum spatial dimension required for such phase transition is 3.9\cite{tarjus}. However, the {\it anisotropy} may have an important role in having such 
phase transition (in three dimensions) in the presence of random field. It has not been studied before. In this paper, this issue is addressed through the Monte Carlo simulation of three dimensional 
anisotropic XY ferromagnet in the presence of external random magnetic field. We have reported our simulational results, mainly,
the dependence of the critical temperature on the magnitude of random field applied. 

We have applied the random field, on the anisotropic XY ferromagnet, in two ways: (i) A-type: the magnitude of the field is fixed and its direction is random between 0 and $2\pi$, with full circular
symmetry and (ii) B-type: the magnitude of random field is fixed but it chooses the direction randomly within angular window. 

In the first case (A-type), we have noticed the reduction of the critical temperature as the magnitude of the applied random field is increased. On the other hand, in the absence of any random field, the anisotropy increases the critical temperature. So, a competition has been observed between anisotropy and the random field. Can one expect any pair of values of the 
strength of anisotropy and the magnitude of random field which preserves the value of the critical temperature\cite{campostrini,hasenbusch} of isotropic XY ferromagnet in three dimensions? Our results provide the answer. The amount of field
(depending on the value of anisotropy) which preserves the critical temperature for isotropic XY ferromagnet in three dimensions may be called the compensating field. We have studied this
functional dependence of the compensating field on the anisotropy.
This dependence is found to be linear.

In the second case (random field in angular window, B-type), the dependence of the critical temperature (for fixed anisotropy and the magnitude of random field) on the angle, is studied. It is observed that the critical temperature decreases as the angle
increases. The result is obvious, for small angular window, the
effect of the random field is not so prominent for the reduction of the critical temperature. The random field having full circular symmetry strongly affects the system and reduces the critical temperature.

All these critical behaviours mentioned above have been formalized by finite size effects to achieve the merit of true thermodynamic phase transitions. The critical exponents for the
susceptibility ($\chi_p \sim L^{{\gamma'} \over {\nu}}$) and that for the
specific heat ($C \sim L^{{\alpha} \over {\nu}}$) are estimated
$1.46\pm0.14$ (with $\chi^2=4.39$ DOF=3) and $0.17\pm0.02$ (with $\chi^2=0.007$ DOF=3) respectively for the case of full circular symmetry (A-type) of the applied random field. For applied random field in an angular window (B-type), we have also estimated, 
${{\gamma'} \over {\nu}}=1.47\pm0.10$ (with $\chi^2=2.36$ DOF=3) and ${{\alpha} \over {\nu}}=0.16\pm0.01$(with $\chi^2=0.007$ DOF=3). 

To the best of  knowledge of these authors, study of this sort has never been conducted before.
The effects of random field on the critical behaviours of anisotropic XY ferromagnet in three dimensions provide many interesting results. 
Moreover, the present study opens up some important ideas for future investigations. It would be interesting to study the effects of random fields where both magnitude and directions are random. The behaviours for single site anisotropy may be a good candidate for further investigations. A similar study can be extended to the classical anisotropic Heisenberg ferromagnet.  We believe that the role of quenched disorder (in the form of random fields) on the critical behaviours of continuous symmetric ferromagnetic models will be an interesting field of research in near future.

It may be worthmentioning here that we have estimated the exponent ${{\beta} \over {\nu}} =0.271$ and 
${{\gamma'} \over {\nu}} = 1.751$ in our previous work \cite{olivia} in the phase transitions of anisotropix XY ferromagnets in three dimensions using Monte Carlo simulation. These exponents neither show the Ising  Universality class nor the XY
universality class in three dimensions. The very strong anisotropy strength would compel the spins to take the discrete orientaions like Ising spins, however, for the small values of anisotropy the spins can have both x and y components which do not lead to entirely Ising symmetry.  This is the main reason for not having  Ising universality class in the presence of small anisotropy we have used in this study. Moreover, the effects of bilinear exchange anisotropy and single site anisotropy are different in the context of spin orientations. Indeed, a huge computational effort is required to have precise determination of the universality class (via precise estimation of the critical exponents), which is beyond the scope of our present investigations. It may be noted here, for this bilinear exchange kind of anisotropy ($\gamma$), the $\gamma=1$ leads the system to the XX model still retaining the continuous symmetry, which is different from discrete Ising symmetry.
\vskip 0.5cm

\noindent {\bf Acknowledgements:}  OM acknowledges
MANF,UGC, Govt. of India for financial support. MA acknowledges FRPDF grant provided by Presidency University, Kolkata, India. We thank Ishita Tikader for a careful reading
of the manuscript.

\vskip 0.2cm

\noindent {\bf Data availability statement:} Data will be available on request to Olivia Mallick.

\vskip 0.2cm

\noindent {\bf Conflict of interest statement:} We declare that this manuscript is free from any conflict of
interest. The authors have no financial or proprietary interests in any material discussed in this
article.

\vskip 0.2cm

\noindent {\bf Funding statement:} No funding was received particularly to support this work.

\vskip 0.2cm

\noindent {\bf Authors’ contributions:} Olivia Mallick-developed the code, collected the data, prepared the
figures, analysed the results, wrote the manuscript. Muktish Acharyya-conceptualized the problem, developed the code,
analysed the results, wrote the manuscript.

\vskip 0.2cm

\noindent {\bf Note added in proof:} Recently, we have noticed an article \cite{rajiv} which reports a perturbative study of XY model with quenched random fields on a fully connected graph. A spin cone has been found as an ordered state. The order-disorder transition has been found to be first order with the variation of the strength of random field at $T=0$.


\newpage

\newpage

\newpage

\newpage

\newpage

\newpage

\newpage

\newpage

\newpage

\newpage

\newpage

\newpage

\begin{figure}[h]
\begin{center}

\resizebox{8cm}{!}{\includegraphics[angle=270]{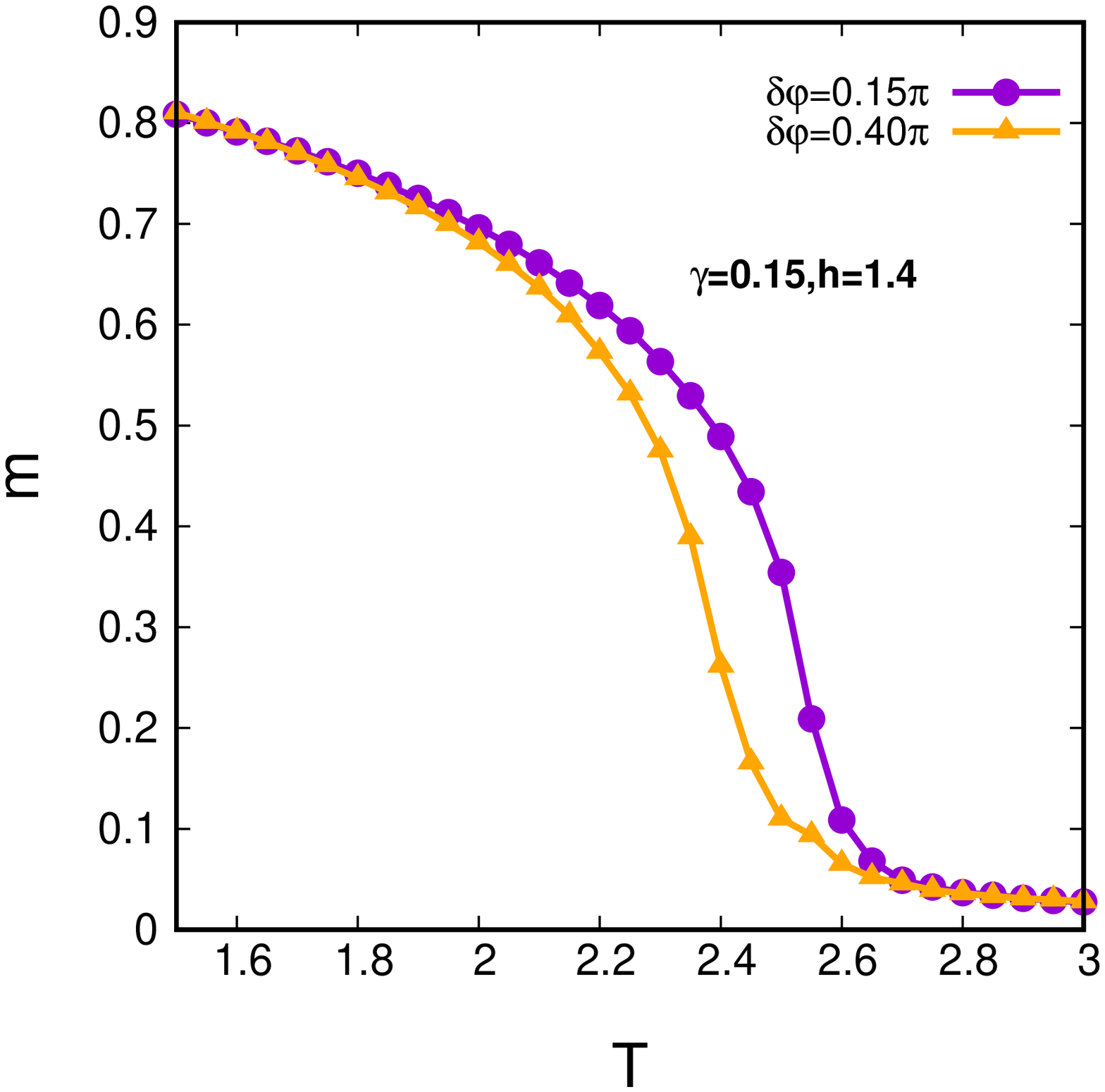}}(a)
\resizebox{8cm}{!}{\includegraphics[angle=270]{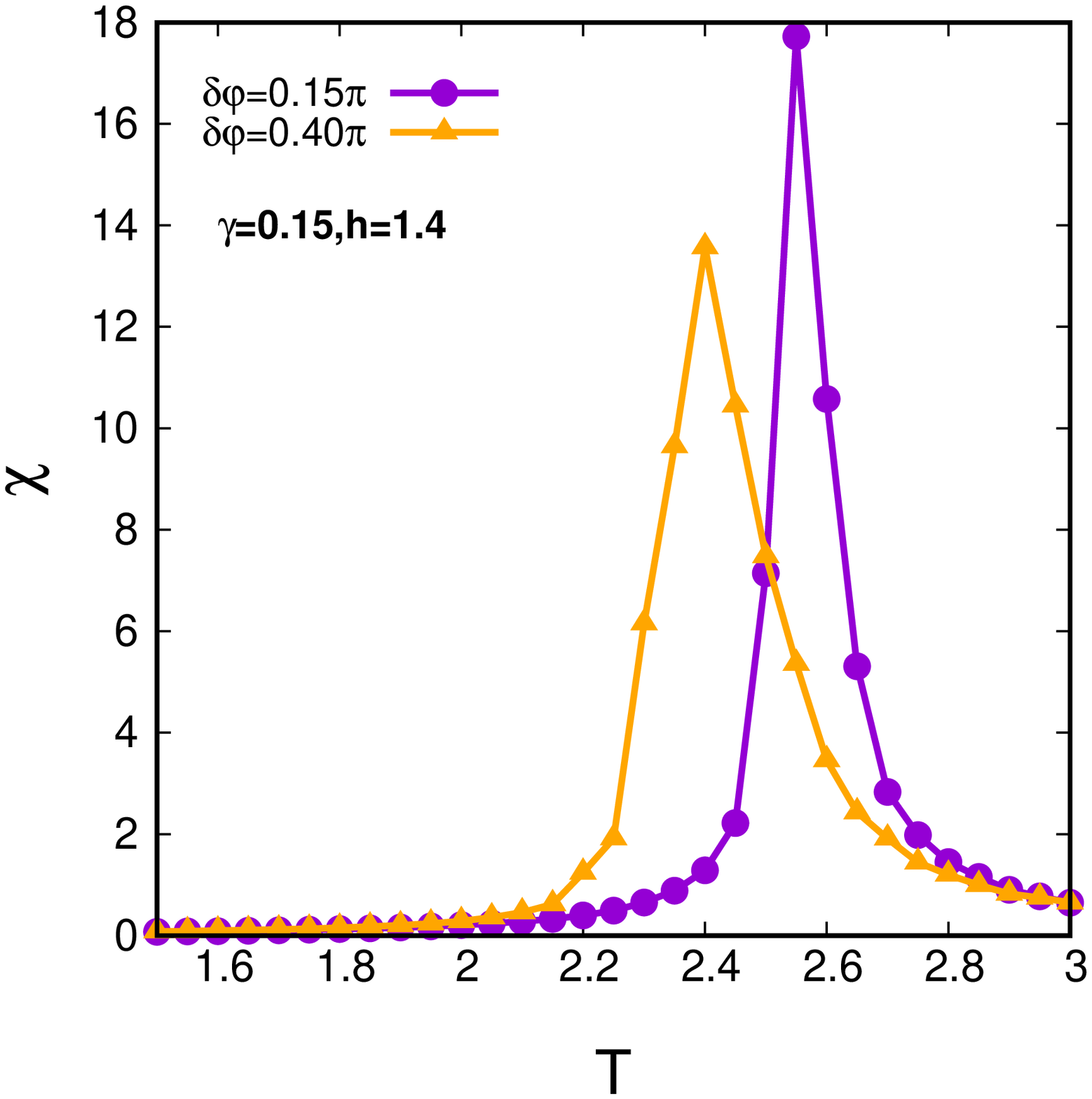}}(b)
\resizebox{8cm}{!}{\includegraphics[angle=270]{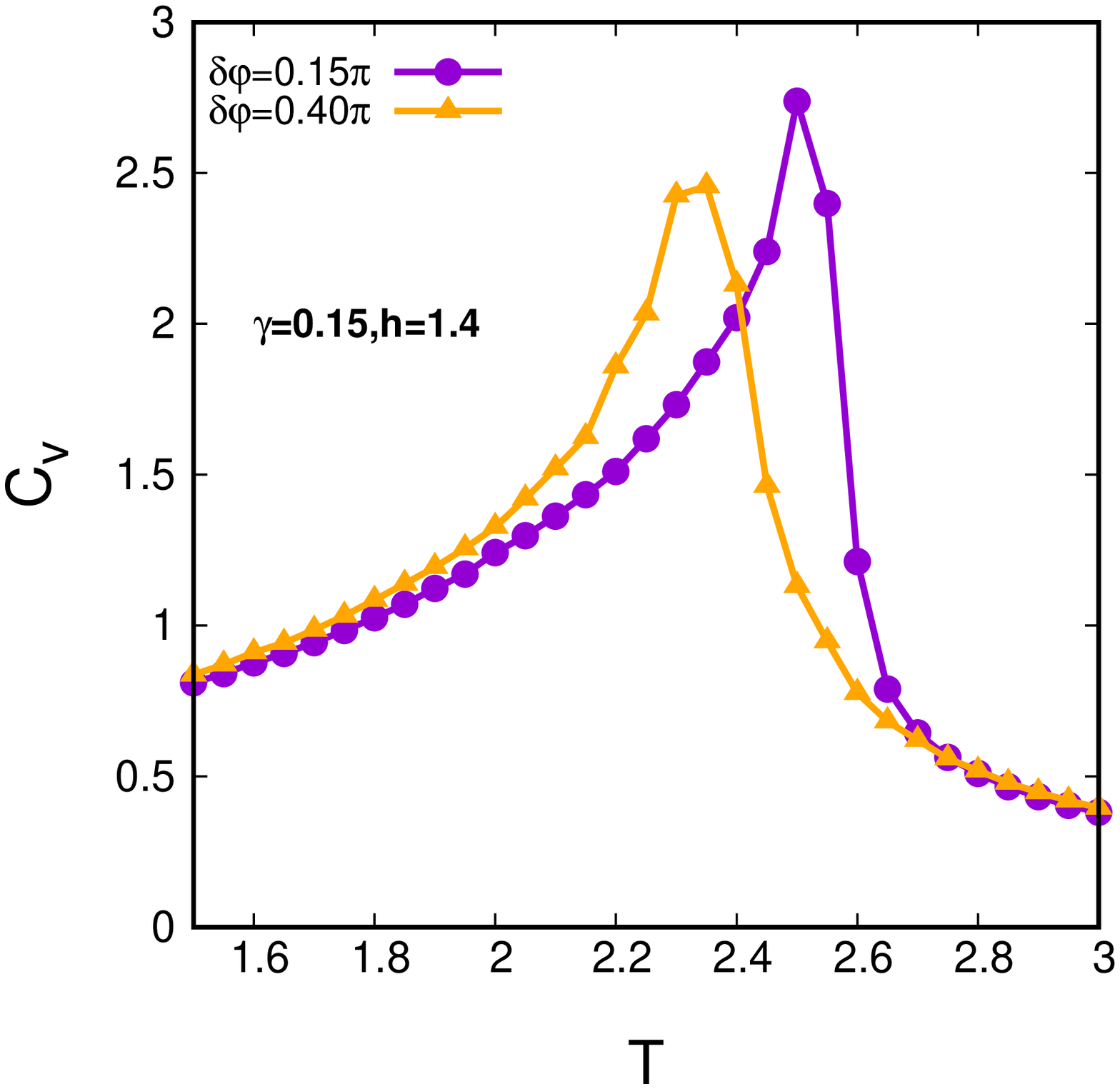}}(c)

\caption{ The (a) magnetization ($m$), (b) susceptibility ($\chi$) and
(c) specific-heat ($C_v$) are plotted against temperature for two different angular window ($\delta\phi=0.15\pi$ and $\delta\phi=0.40\pi$) with anisotropy $\gamma=0.15$ and field strength $h=1.4$.}

\label{m-conical-h.eps}
\end{center}
\end{figure}

\newpage
\begin{figure}[h]
\begin{center}

\resizebox{15cm}{!}{\includegraphics[angle=270]{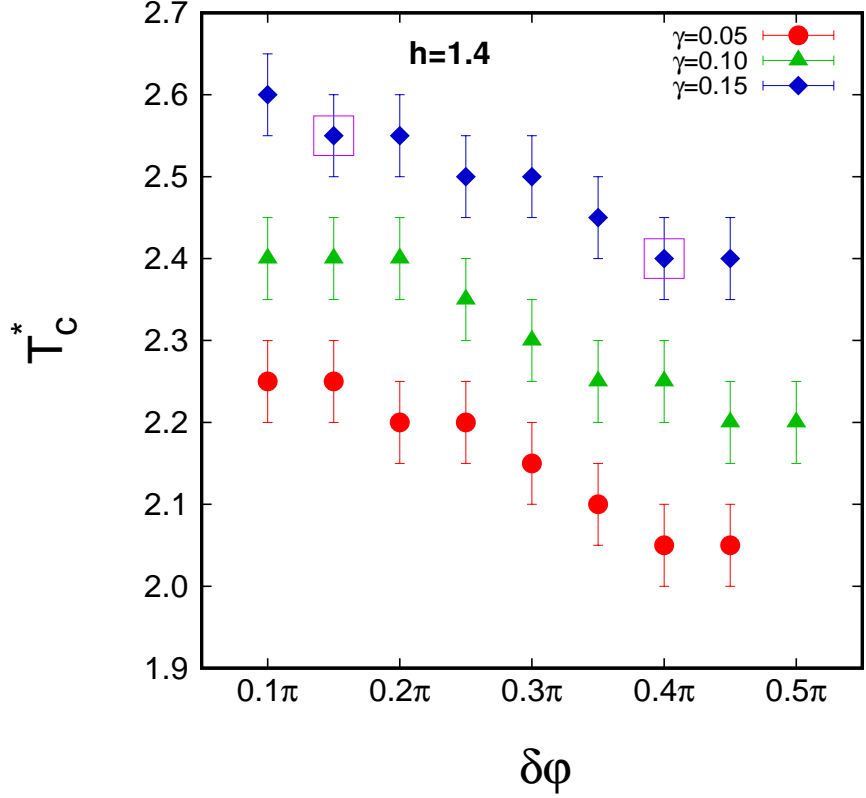}}

\caption{The variation of the pseudocritical temperature ($T^*_c$) with the angular extension ($\delta\Phi$) of the
random field for the magnitude  $h=1.4$. The three different symbols represents three different  strengths
of the anisotropy ($\gamma$). The $4\delta\Phi=2\pi$ restores the full circular symmetry of the random field. The pseudocritical
temperature for $\gamma=0.1$ and $h=1.4$ for circularly symmetric ($\delta\Phi={{\pi} \over {2}}$) random field is shown by
marking the data point. The data point (bounded by the boxes) represent the pseudocritical temperatures (obtained
from the peak position of the susceptibility) for two different values of $\delta\Phi$ ($0.15\pi$ and $0.40\pi$) with fixed
anisotropy $\gamma=0.15$ and $h=1.4$ (see Fig-\ref{m-conical-h.eps}). The errorbars are the maximum range arises from the stepsize of cooling
the system.}

\label {ang-Tc.eps}
\end{center}
\end{figure}

\newpage

\newpage

\newpage

\newpage

\newpage


\end{document}